\documentclass[twocolumn]{aastex631}
\usepackage[utf8]{inputenc}
\usepackage{multirow}

\shorttitle{Control sample for TOI host stars}
\shortauthors{Littlefield et al.}

\begin{document}

\title{High-Resolution Imaging of a TESS Control Sample: Verifying a Deficit of Close-In Stellar Companions to Exoplanet Host Stars}

\author[0000-0001-7746-5795]{Colin Littlefield}
\affiliation{Bay Area Environmental Research Institute, Moffett Field, CA 94035, USA}

\author[0000-0002-2532-2853]{Steve~B.~Howell}
\affil{NASA Ames Research Center, Moffett Field, CA 94035, USA}

\author[0000-0002-5741-3047]{David R. Ciardi}
\affiliation{NASA Exoplanet Science Institute, Caltech/IPAC, Mail Code 100-22, 1200 E. California Blvd., Pasadena, CA 91125, USA}
\author[0000-0002-9903-9911]{Kathryn V. Lester}
\affil{NASA Ames Research Center, Moffett Field, CA 94035, USA}

\author[0000-0002-0885-7215]{Mark~E.~Everett}
\affiliation{NSF’s National Optical-Infrared Astronomy Research Laboratory, 950 N. Cherry Ave., Tucson, AZ 85719, USA}

\author[0000-0001-9800-6248]{Elise Furlan}
\affiliation{NASA Exoplanet Science Institute, Caltech/IPAC, Mail Code 100-22, 1200 E. California Blvd., Pasadena, CA 91125, USA}

\author[0000-0001-7233-7508]{Rachel A. Matson}
\affiliation{U.S. Naval Observatory, 3450 Massachusetts Avenue NW, Washington, D.C. 20392, USA}

\author[0000-0001-9309-0102]{Sergio B. Fajardo-Acosta}
\affiliation{Caltech/IPAC, Mail Code 100-22, 1200 E. California Blvd., Pasadena, CA 91125, USA}

\author[0000-0003-2519-6161]{Crystal~L.~Gnilka}
\affil{NASA Ames Research Center, Moffett Field, CA 94035, USA}

\begin{abstract}

    The large number of exoplanets discovered with the Transiting Exoplanet Survey Satellite (TESS) means that any observational biases from TESS could influence the derived stellar multiplicity statistics of exoplanet host stars. To investigate this problem, we obtained speckle interferometry of 207 control stars whose properties in the TESS Input Catalog (TIC) closely match those of an exoplanetary host star in the TESS Object of Interest (TOI) catalog, with the objective of measuring the fraction of these stars that have companions within $\sim1.2"$. Our main result is the identification of a bias in the creation of the control sample that prevents the selection of binaries with $0.1" \lesssim \rho \lesssim 1.2"$ and $\Delta$mag $\lesssim3$. This bias is the result of large astrometric residuals that cause binaries with these parameters to fail the quality checks used to create the TIC, which in turn causes them to have incomplete stellar parameters (and uncertainties) in the TIC. Any stellar multiplicity study that relies exclusively upon TIC stellar parameters to identify its targets will struggle to select unresolved binaries in this parameter space. Left uncorrected, this selection bias disproportionately excludes high-mass-ratio binaries, causing the mass-ratio distribution of the companions to deviate significantly from the uniform distribution expected of FGK-type field binaries. After accounting for this bias, the companion rate of the FGK control stars is consistent with the canonical 46$\pm$2\% rate from \citet{raghavan}, and the mass-ratio distribution agrees with that of binary TOI host stars. There is marginal evidence that the control-star companions have smaller projected orbital separations than TOI host stars from previous studies.

\end{abstract}

\section{Introduction}

    The ever-growing tally of exoplanets discovered by the Transiting Exoplanet Survey Satellite \cite[TESS;][]{ricker} has led to great interest in these planets' host stars, particularly those that harbor stellar companions. Even beyond its intrinsic astrophysical implications for planet formation, the presence of a companion star can result in a number of biases, ranging from the systematic underestimation of planetary radii \citep{ciardi15} to the miscalculation of exoplanet occurrence rates \citep{hirsch, bouma18}. 
    
    The presence of a stellar companion can profoundly influence the process of planet formation. For example, \citet{kraus} found that the presence of a close ($\lesssim40$~AU) stellar companion causes protoplanetary disks to be dispersed much faster ($\lesssim1$~Myr) than in wide binaries or single stars ($1-3$~Myr), leading to a deficit of planets in those systems in comparison to single stars. And in a high-resolution-imaging study of 382 Kepler Objects of Interest (KOIs), \citet{kraus16} determined that planets are underabundant by a factor of $\sim3$ in binaries with separations $\lesssim50$~AU (at $4.6\sigma$ significance). At wider separations, \citet{kraus16} found little evidence of suppression, a result somewhat at odds with \citet{wang14}, who concluded ( with 1-2$\sigma$ significance) that the suppression of circumstellar planets extends to stellar binary separations as wide as $\sim$1500~AU for Kepler based on adaptive-optics imaging and radial velocity analysis of a sample of KOIs. Despite this mild disagreement about planet suppression at large binary separations, both \citet{wang14} and \citet{kraus16} concur that planet suppression is most pronounced at small binary separations. Further, \citet{hirsch21} found that the occurrence rate of giant planets in widely separated ($100$~AU) binaries was identical to that of single stars, while in close binaries ($<100$ AU), it was reduced by a factor of $4-5$. In line with this picture, a series of studies, including \citet{ziegler20}, \citet{howell21}, and \citet{lester21}, used high-contrast imaging to find that exoplanet-hosting TESS Objects of Interest (TOIs)\footnote{See \citet{guerrero} for a description of the process used to identify TOIs.} show a deficit of stellar companions within 100~AU in comparison to field binaries.

    As work progresses on TOI host stars, a subtle question arises: to what extent might the differences with respect to field binaries be attributable to observational biases? \citet{ziegler20}, \citet{howell21}, and \citet{lester21} compared the properties of binary TOI host stars against the properties of FGK dwarfs from \citet{raghavan}. Although a volume-limited study like \citet{raghavan} should combat observational biases in order to yield statistically robust results for stellar populations, \citet{stassun18} pointed out that the main scientific objective of TESS is to maximize the number of small planets detected via the transit method---and not to yield a statistically unbiased representation of the population of extrasolar planetary systems. It is therefore unclear whether binary statistics derived from \citet{raghavan} provide the best point of comparison for TESS exoplanet hosts.

    Comparing the binary rates of transiting planet host stars to the solar neighborhood sample from \citet{raghavan} makes the implicit assumption that the there are no systematic differences between the TOI sample and the volume limited solar neighborhood sample.  For example, the planet content of the \citet{raghavan} sample is not well characterized.  Toward the goal of trying to compare the binarity rate of TOIs to that of the general field, we have constructed a control sample of stars observed by TESS but not found to host transiting exoplanets. The planetary abundance of these control stars is unknown but it is known that no close-in transiting planets have been detected around them. Our aim was to use this control sample to see whether its binarity rate was systematically different than the binarity rate of stars found to host close-in transiting planets, as seen by the same instrument (TESS).

    In the process of conducting this experiment, we discovered a bias in the TESS Input Catalog whereby well-characterized stars (e.g., stars with complete and reliable stellar parameters) are preferentially single stars. In this paper, we describe the selection of the control sample and the discovery of the bias. We have worked to correct for this bias and make a true assessment of the binarity rate of FGK stars that were observed with TESS and found to not host a close-in transiting planet.

\section{Data}

    \subsection{Selection of control stars}
    
        Our objective was to pair individual TOIs with a non-TOI control star of comparable $T_{eff}$, radius, and distance. We excluded from consideration TOIs with a TFOPWG disposition of ``FP'' (false positive).  For each of the remaining TOIs, we searched version 8.2 \citep{paegert} of the TESS Input Catalog (TIC) for stars with similar  effective temperature ($T_{eff}$), radius, and distance relative to each specific TOI.\footnote{The distance requirement is particularly important for high-contrast imaging because it ensures identical spatial resolution for both a TOI and its control star.} This left us with a large, preliminary list of candidate control stars for each TOI. To identify the closest matches for any given TOI, we added in quadrature the fractional differences in $T_{eff}$, stellar radius, and distance between that particular TOI and the candidate control star. The resulting metric enabled us to create a ranked list of suitable control stars for each TOI.

        We eliminated any control stars that were candidate exoplanet hosts, and we also imposed a requirement that they have been observed by TESS in at least one sector. This restriction provided some confidence that the control stars lack detectable transiting exoplanets. The non-detection of candidate exoplanets by TESS does not establish that these stars lack exoplanetary systems; TESS can detect exoplanets only if (1) their orbital inclinations are sufficiently high to result in a transit of the host star, (2) the transits occur during the TESS observing window, and (3) the transit depth is sufficiently deep to be detected by the analysis pipeline. However, our goal was to compare TOI stars against otherwise identical non-TOI stars.

        The TIC v.8.2 stellar parameters are foundational to our methodology, so it is worthwhile to briefly summarize their provenance, which \citet{stassun19} explain in meticulous detail. The TIC v.8.2 computes stellar parameters for stars with good Gaia DR2 astrometry and photometry. For the stars that survive this initial cut, $T_{eff}$ is systematically calculated from the dereddened {\it Gaia} DR2 colors of each star except in special cases where a spectroscopic $T_{eff}$ is available. Given the $T_{eff}$ and the {\it Gaia} DR2 distance, the stellar radius is calculated by applying the Stefan-Boltzmann relation. \citet{stassun19} caution that these calculations make no allowance for the reality that some apparently single stars are actually unresolved binaries and that the presence of a companion can cause the computed stellar parameters to be unreliable. For example, if an unresolved binary contained two identical stars, the stellar radius could be overestimated by a factor of $\sqrt{2}$. Additionally, for some specially curated lists of targets, the stellar parameters therein supersede the parameters computed by the TIC pipeline.

        \subsubsection{Comparison of control stars to TOIs}
        
\begin{figure*}
            \centering
            \includegraphics[width=\textwidth]{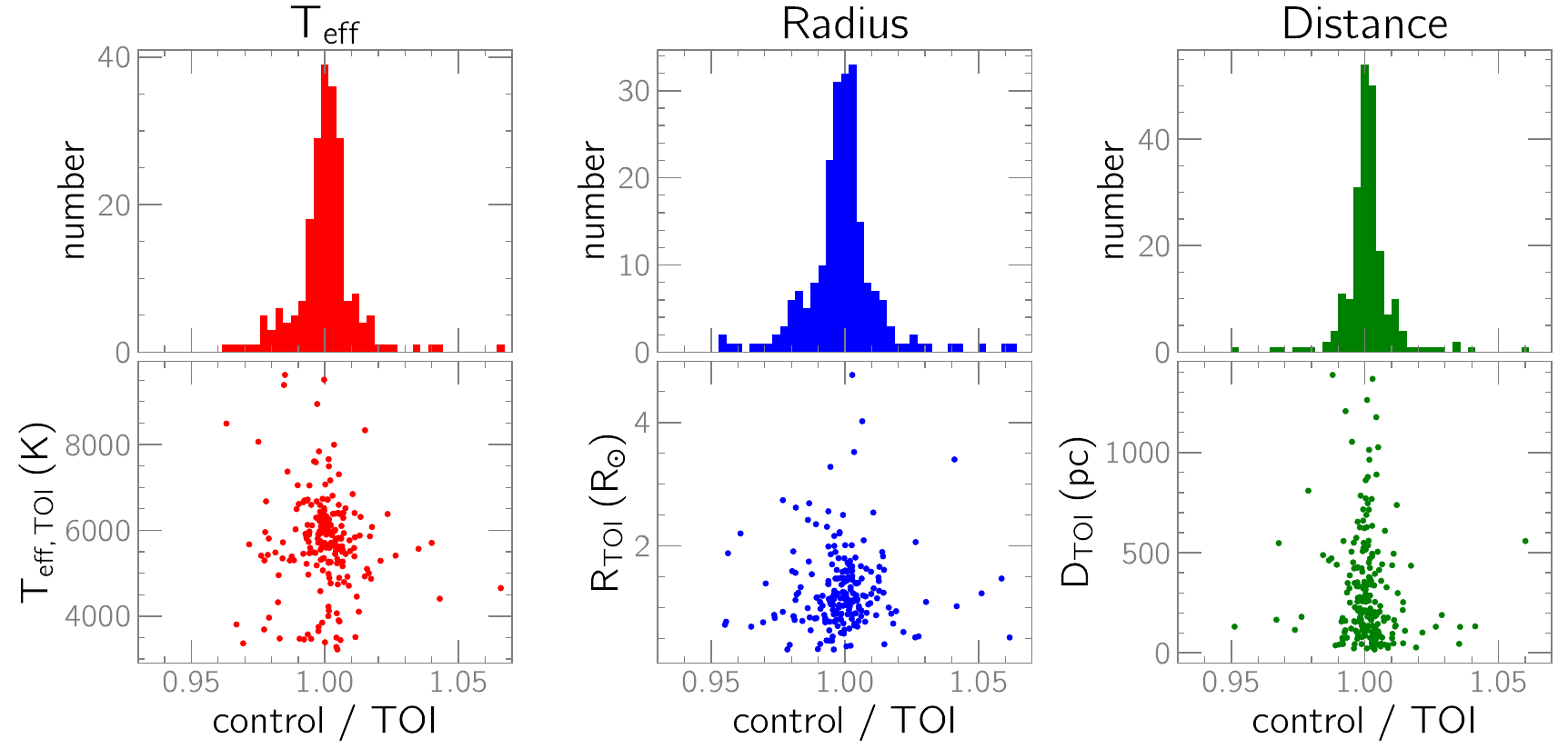}
            \caption{Comparison of effective temperature, stellar radius, and distance between each observed control star and its corresponding TOI. The abscissa in each panel is the fractional difference between a control star's  property and that of the TOI.  
            }
            \label{fig:match_statistics}
        \end{figure*}

        The histograms in Figure~\ref{fig:match_statistics} confirm that the observed control stars closely match the properties of their corresponding TOI stars, with only small fractional differences in the effective temperature, radius, and distance. The control stars span a wide range of effective temperature along the main sequence, as established in Figure~\ref{fig:color-magnitude}.

        We also investigated whether the observed TESS magnitudes of the control stars agree with those of the TOIs. While this was not explicitly controlled for in our selection, we would expect the magnitude differences to be similar because in the absence of interstellar extinction, the stellar effective temperature, radius, and distance should determine, via the Stefan-Boltzmann relation, the apparent magnitude. 
        
        The resulting distribution is symmetric and centered on 0, with over 90\% of the differences being smaller than 0.2 mag, which is reassuring. Nevertheless, there were four extreme outliers where the TOI and control stars had magnitude differences greater than 1 magnitude. We used the \citet{green19} three-dimensional interstellar reddening maps to investigate these outliers, and in each case, there were significant differences in the interstellar extinction along the two lines of sight. Although the computation of the TESS magnitude does apply a correction for extinction \citep{stassun19}, we suspect that the four TOI-control star pairings with large magnitude differences might be the result of an imperfect dereddening that propagated into the computations of the TESS magnitude, $T_{eff}$, and the stellar radius.

    \begin{figure*}
            \centering
            \includegraphics[width=\textwidth]{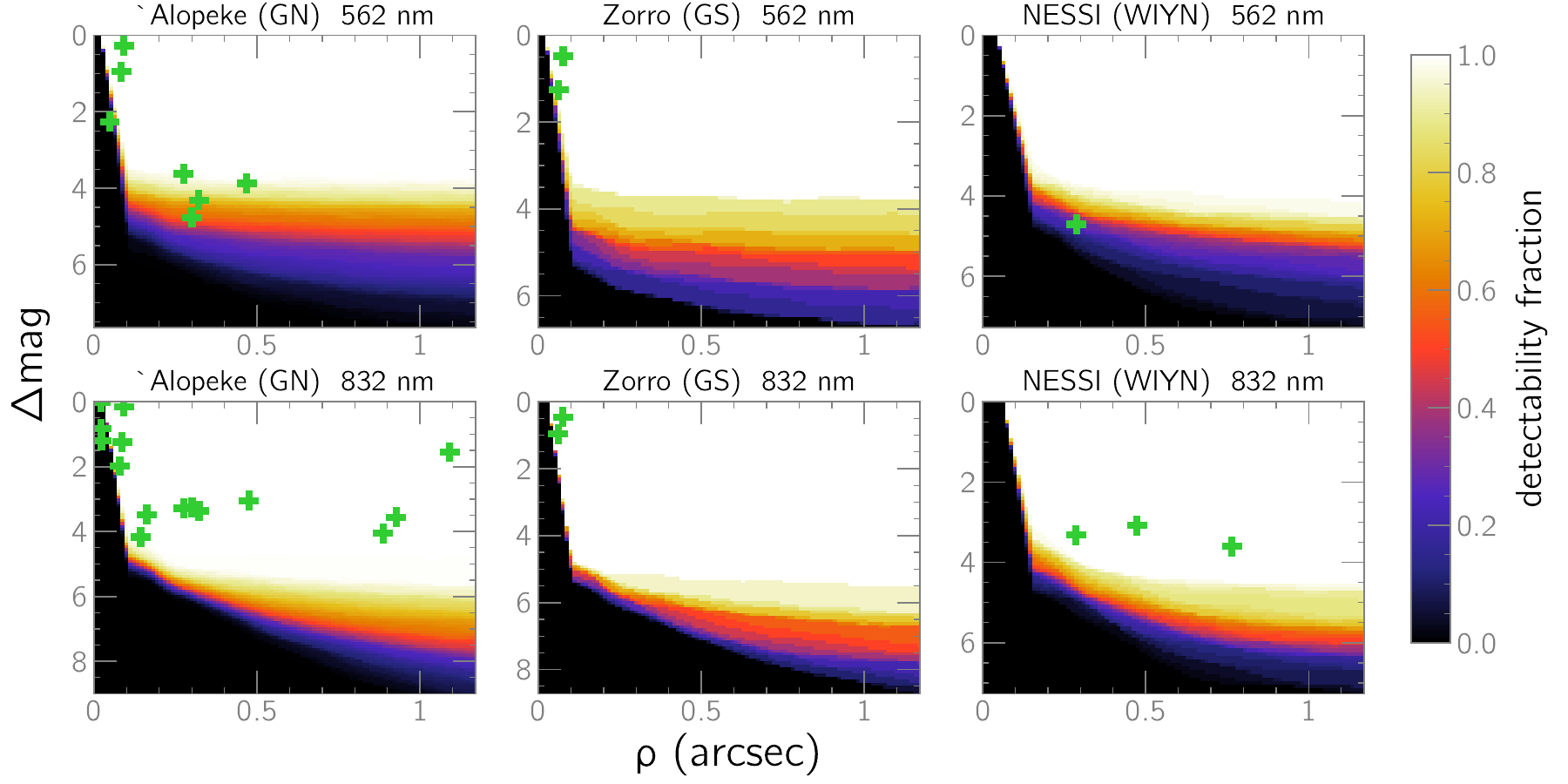}
            \caption{Two-dimensional histograms of the contrast curves used in this study, separated by instrument and bandpass. Detected companions within $\rho < 1.2"$ are marked with green crosses. Companions detected at wider separations (six at 832~nm and two at 562~nm) are not shown. Many of the companions detected at 832~nm were undetected at 562~nm. The color scale represents the fraction of contrast curves in which a companion could have been detected at a given combination of $\rho$ and $\Delta$mag. Note that the range of $\Delta$mag varies between the panels because the contrast curves vary from from instrument to instrument.
            }
            \label{fig:contrastcurves}
    \end{figure*}

    \subsection{Observational technique}
    
        Our observations were carried out with the `Alopeke and Zorro speckle cameras \citep{scott21} on the 8-meter Gemini North and Gemini South telescopes, respectively. `Alopeke and Zorro are identical instruments. Each contains a dichroic, positioned at a $45^{\circ}$ angle relative to the collimated input beam, to split the beam into blue ($\lambda < 674$~nm) and red ($\lambda > 674$~nm) components. Each beam travels to its own dedicated electron-multiplying CCD camera, so both `Alopeke and Zorro obtain simultaneous dual-band observations with a field of view of $2.5"\times2.5"$. To boost the contrast of the speckles, our observations utilized narrow-band filters, centered at 562~nm (44~nm FWHM) and 832~nm (40~nm FWHM) for the blue and red cameras, respectively. 
        
        Speckle observations with these instruments were obtained in sets of 1000 consecutive frames, each with an exposure time of 60~milliseconds with negligible overhead. Depending on the brightness of the target and the sky conditions, we typically obtained anywhere from 3 to 15 sets. Some of our observations were obtained using the 3.5-m  WIYN telescope and the NN-EXPLORE Exoplanet Stellar Speckle Imager \citep[NESSI;][]{scott_nessi} instrument, which is conceptually similar to (and served as the prototype for) both `Alopeke and Zorro. The NESSI observations used the same filters as the Gemini observations, but the resolution is worse by a factor of $\sim2$ because of the smaller aperture.

        Within a short time of each observation of a control star, we obtained speckle observations of a bright, single star within a few degrees of the control star in order to measure the speckle transfer function. We reduced the data to obtain high-level data products using a pipeline described in \citet{howell_pipeline}. For any detected companion stars, we measured the angular separation and position angle relative to the primary target along with the magnitude difference.  Another pipeline product is a reconstructed image of the focal plane centered on the target. From this, we measured a contrast curve that represents the flux corresponding to a 5$\sigma$ deviation from the mean within each of a set of concentric annuli centered on the target star. This contrast curve is smoothed slightly to remove deviations caused by known artifacts or the flux from a detected companion. The contrast curve is estimated out to a maximum separation of 1.2", a limit that avoids problems caused by potential centering errors of the target within the readout section of the detectors and the loss of photometric sensitivity at wider separations where speckle patterns can start to become uncorrelated. `Alopeke (Gemini North) and Zorro (Gemini South) generally achieve better contrast than NESSI, and for all instruments, the 832~nm filter tends to have deeper contrast limits than the simultaneous 562~nm data.

        To measure the pixel scale and the position angle of the detectors, we periodically observe known binary stars whose orbital elements are listed in an updated version of the \citet{hartkopf} catalog.\footnote{\url{https://crf.usno.navy.mil/wds-orb6}.} The pixel scale and array orientation are calculated for each observing run, and while they tend to be quite stable between observing runs, we updated them as needed.

\section{Results}

    \subsection{Companions to Control Stars }
     
        We observed a total of 207 control stars, of which 26 had detected binary companions at 832~nm and 12 had companions at 562~nm (Figure~\ref{fig:contrastcurves}). Of these companions, 20 had angular separations of $\rho<1.2"$ at 832~nm with respect to the primary star, while 10 of the companions detected at 562~nm satisfied $\rho<1.2"$. 

        \begin{deluxetable*}{c|ccc|ccc}
            \label{table:companions}
            \tablecaption{Properties of detected companions to control stars}
            \tablehead{
                \colhead{TIC ID} & 
                \colhead{$\rho$ (arcsec)} & 
                \colhead{$\Delta\mathrm{mag}$} &
                \colhead{$\theta$ (deg)} & 
                \colhead{$\rho$ (arcsec)} & 
                \colhead{$\Delta\mathrm{mag}$} &
                \colhead{$\theta$ (deg)}  \\
                \colhead{} &
                \colhead{(832 nm)} &
                \colhead{(832 nm)} &
                \colhead{(832 nm)} &
                \colhead{(562 nm)} &
                \colhead{(562 nm)} &
                \colhead{(562 nm)} 
                }
                    \startdata
                    372086900 & 0.021 & 0.02 & 208.19 &    &    &    \\
                    377058463 & 0.022 & 0.83 & 241.739 & 0.048 & 2.26 & 240.839 \\
                    196383895 & 0.023 & 1.2 & 296.698 &    &    &    \\
                    296781193 & 0.061 & 0.96 & 183.573 & 0.062 & 1.26 & 183.993 \\
                    437327600 & 0.075 & 0.46 & 202.852 & 0.076 & 0.48 & 202.663 \\
                    14899687 & 0.079 & 1.98 & 265.021 &    &    &    \\
                    301482228 & 0.086 & 1.24 & 63.579 & 0.083 & 0.95 & 65.537 \\
                    267686220 & 0.091 & 0.15 & 243.797 & 0.091 & 0.27 & 243.622 \\
                    264485499 & 0.143 & 4.17 & 358.335 &    &    &    \\
                    355691670 & 0.162 & 3.48 & 335.28 &    &    &    \\
                    85274754 & 0.275 & 3.28 & 125.126 & 0.275 & 3.62 & 125.507 \\
                    354442089 & 0.286 & 3.32 & 232.234 & 0.288 & 4.71 & 233.203 \\
                    194461161 & 0.301 & 3.25 & 102.956 & 0.301 & 4.77 & 102.512 \\
                    91277756 & 0.322 & 3.36 & 18.95 & 0.321 & 4.31 & 18.496 \\
                    376688975 & 0.473 & 3.08 & 22.114 &    &    &    \\
                    269390255 & 0.475 & 3.05 & 149.225 & 0.468 & 3.87 & 148.478 \\
                    83958546 & 0.764 & 3.6 & 245.005 &    &    &    \\
                    301482610 & 0.887 & 4.05 & 338.992 &    &    &    \\
                    241257501 & 0.927 & 3.56 & 327.102 &    &    &    \\
                    96876685 & 1.091 & 1.55 & 203.39 &    &    &    \\
                    140343515 & 1.218 & 5.13 & 317.466 &    &    &    \\
                    716026635 & 1.312 & 3.73 & 314.706 &    &    &    \\
                    91597865 & 1.319 & 2.81 & 60.841 & 1.331 & 2.8 & 62.032 \\
                    2100594 & 1.689 & 9.99 & 323.692 &    &    &    \\
                    306125356 & 1.725 & 5.0 & 70.136 &    &    &    \\
                    411551642 & 2.101 & 2.35 & 353.914 & 2.085 & 3.61 & 354.409 \\
                    \enddata
        \end{deluxetable*}

        To assess whether any of the apparent companions might be unrelated stars along the line of sight to the primary, we retrieved a list of all cataloged TIC v.8.2 stars within a radius of 15~arcmin centered on each control star with a detected companion; built upon Gaia DR2, the TIC v.8.2 is complete to well below the limiting magnitude of our observations. Within this search region, we counted the number of stars within $\pm$1~mag of the detected companion. This yielded a localized estimate of the number density of sources of comparable brightness to the putative companion, which lets us estimate the probability of a chance alignment that causes a physically unrelated field star to masquerade as an apparent companion. We found that there is very little chance that any of the detected companions are physically unrelated alignments. In some cases, there were no field stars of comparable brightness to the putative companion; at the opposite extreme, the control-star companion in the densest star field had a line-of-sight probability of just $0.2\%$. We conclude that all 26 companions are probably physically associated with their control stars, a finding that is consistent with previous analyses about the likelihood of boundedness of close-in companions \citep[e.g., ][]{everett15, hirsch}.

        Table~\ref{table:companions} lists the properties of the detected companions, while Table~\ref{table:control_sample} provides the correspondence between the control stars and TOIs.
    
    \subsection{Simulations of Binary Companions} \label{sec:simulation}

        It is straightforward to describe the properties of the companions that we observed around the control stars. But for the apparently single control stars, the absence of a detected companion might mean that a control star is single, that any companions are too faint to be detected, or that the binary angular separation might be extremely small or too large to fit inside the speckle field of view. Speckle interferometric studies by \citet{matson19} and \citet{howell21} addressed these issues by calculating three separate fractions. The first, the ``companion fraction,'' estimates the fraction of all plausible companions whose $\Delta$mag would be above the contrast limits (``the companion fraction'').     The second is the fraction of semimajor axes that would cause the companion's angular separation to be within the instrument's field of view (``the distribution fraction''). Finally, the ``speckle fraction'' is the product of the companion and distribution fractions, and it estimates the percentage of companions that would be detectable with speckle observations.

        To calculate the speckle fraction for each star in our control sample, as well as the predicted distributions for the mass ratio and the binary separations, we use a Monte Carlo simulation similar to the ones designed by \citet{ziegler20} and \citet{lester21}. For each control star, we retrieve its mass and distance, as well as their associated uncertainties, from the TIC. Each iteration of the simulation then proceeds as follows. We add Gaussian noise to both the stellar mass and distance and randomly draw a binary mass ratio $q$ from the \citet{raghavan} distribution, which has a uniform probability for $0.2 \leq$ q $\leq 0.95$, with a doubled probability above q=0.95. We further draw a random orbital period from the \citet{raghavan} log-normal distribution of $\log P = 5.03$~d, $\sigma_{\log P} = 2.28$.
            
        Given the two stellar masses and their orbital period, we calculate their semimajor axis. We then randomly draw from uniform distributions the binary inclination, argument of periastron, longitude of the ascending node, and orbital eccentricity. With these simulated orbital elements, we calculate the angular separation of the synthetic binary companion at a random orbital phase. We also calculate $\Delta$~mag in both the $V$ and TESS bandpasses based on the stellar masses and the Modern Mean Stellar Dwarf Sequence \citep[MMSDS;][]{pm13}.\footnote{Although the MMSDS does not list TESS magnitudes, we compute them from the $G$ and $J$ magnitudes in the MMSDS following the conversion procedure in \citet{stassun18}.}
    
        To determine whether the simulated companion would be detectable with our observations, we retrieve the observed contrast curve for the control star, which shows the 5$\sigma$ limit for $\Delta$mag as a function of the angular separation $\rho$. If the simulated companion is above the contrast curve and within 1.2" of the control star, we classify it as detected; otherwise, we classify it as a non-detection. 
        
        We impose the upper limit of 1.2" because it is the limit of the parameter space examined in the contrast curves; above this separation, speckles can become decorrelated \citep{horch11}. Nevertheless, binaries with wider separations are occasionally detected by our pipeline, which contradicts the assumption in our simulations that companions are undetectable when $\rho > 1.2$~arcsec. 

        Since the FGK binary parameters from \citet{raghavan} cannot be extrapolated to other spectral types \citep[e.g., M-dwarfs;][]{winters19}, we further limit the simulations to those systems with FGK-type primary stars (estimated from their effective temperatures in the TIC). Fifteen companion-harboring systems in our control sample satisfy the spectral-type and angular-separation constraints.

        \begin{figure*}
            \centering
            \includegraphics[width=\textwidth]{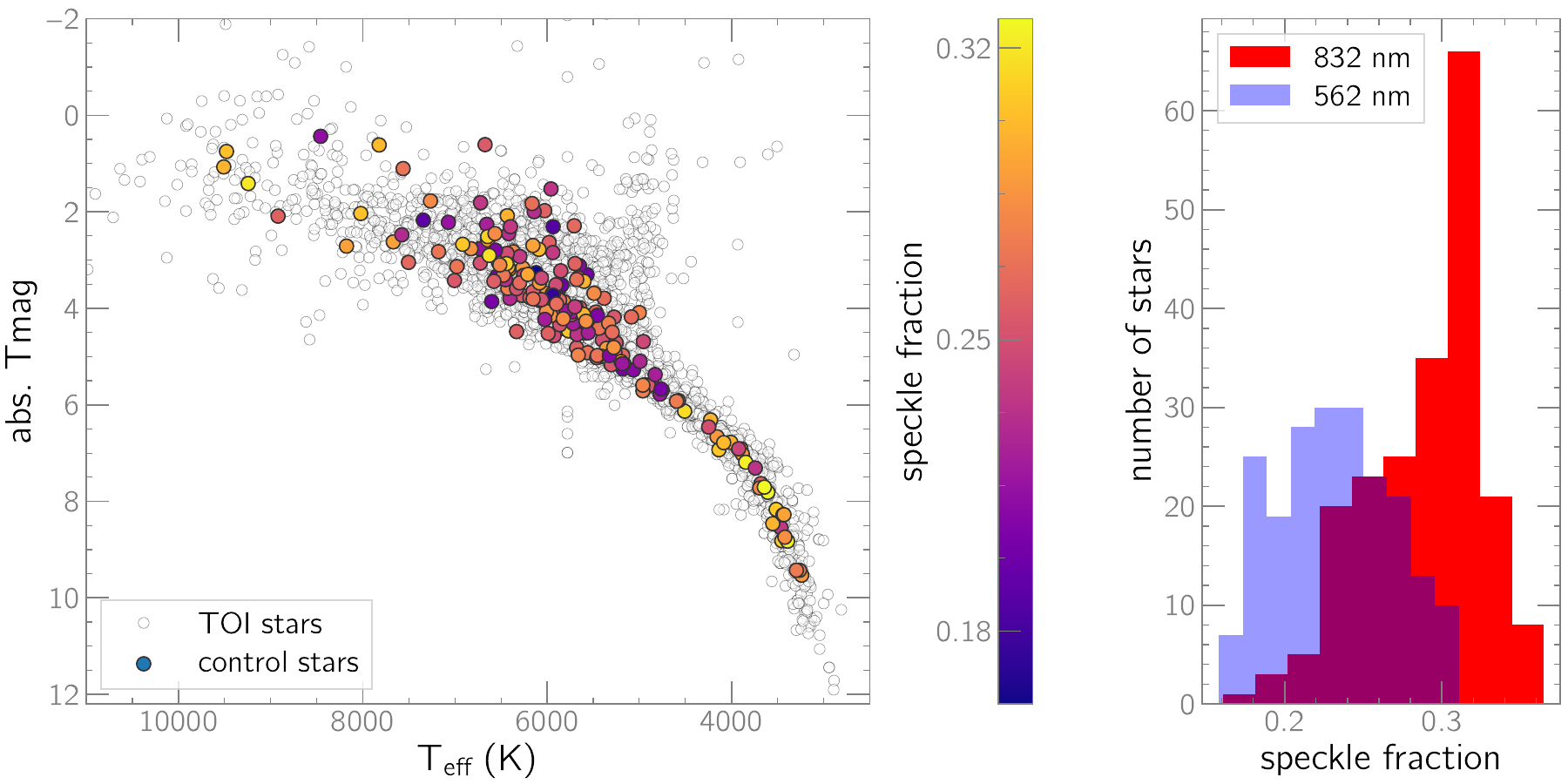}
            \caption{{\bf Left:} Color-magnitude diagram of the control-star sample compared to that of all TOIs (white circles), showing the absolute TESS magnitude as a function of effective temperature. The color of each control-star marker indicates the 832~nm speckle fraction, defined as the fraction of possible companions that could be detected in that bandpass based on our contrast limits, the instrument field of view, and the \citet{raghavan} binary parameters. {\bf Right:} Histograms of the speckle fraction for each bandpass.}
            \label{fig:color-magnitude}
        \end{figure*}

        The speckle fraction for each star is simply the fraction of the 1000 simulations that produced a detectable companion, subject to the aforementioned constraints. Figure~\ref{fig:color-magnitude} shows that the speckle fractions for the control sample are uncorrelated with a control star's location in a color-magnitude diagram, and it further establishes that the speckle fraction tends to be larger at 832~nm than at 562~nm.

        One of the main limitations of our simulations is the uncertainty of the contrast curves at separations within $\rho < 0.1"$. At larger separations, the limiting magnitude is estimated empirically by calculating a 5$\sigma$ threshold from the minima and maxima in the reconstructed speckle image as a function of $\rho$, but at small radii, simple geometric considerations mean that there are relatively few minima and maxima upon which to estimate a detection threshold. Accordingly, $\Delta$~mag is assumed to be 0 at the diffraction limit and is then linearly interpolated to the beginning of the empirical contrast curve at 0.1~arcsec. However, in good seeing conditions, we occasionally detected companions within $\rho < 0.1"$ at values of $\Delta$~mag that are nominally below our detection threshold (Figure~\ref{fig:contrastcurves}), raising the prospect that our contrast curves are too pessimistic about the actual capability of `Alopeke and Zorro to detect companions at very small separations. Crucially, this issue works in one direction: it will cause genuinely detectable combinations of $\rho$ and $\Delta$~mag to be wrongly classified as undetectable in our simulations. As a result, our simulations likely underpredict the number of detectable companions and therefore provide a lower limit.


        The simulations allow us to compute the expected number of detected companions for an assumed orbital-period distribution. To estimate the expected number of detectable companions within 1.2" of FGK stars, we compute the average number of synthetic companions that were detected in the 1,000 simulations. Because there is a companion in each simulation (i.e., a presumed multiplicity rate of 100\%), we then multiply by the FGK dwarf multiplicity rate of $46\pm2$\% from \citet{raghavan} to take into account that slightly over half of the FGK dwarfs are expected to be single. This leaves us with a predicted tally of detected companions.

        \subsection{The absence of close, high-mass-ratio binaries}

            \begin{figure*}
                \centering
                \includegraphics[width=\textwidth]{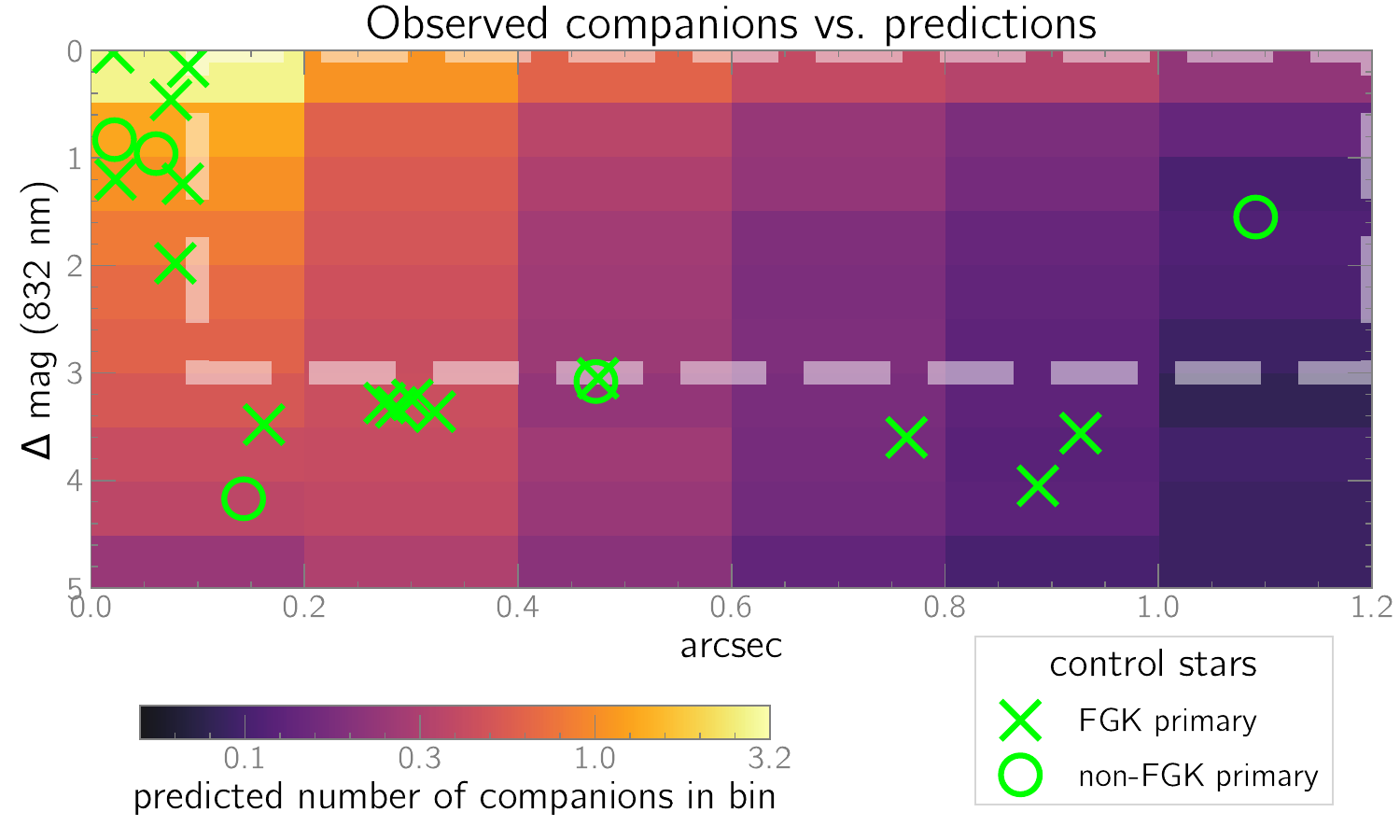}
                \caption{Two-dimensional histogram showing the predicted number of companions for FGK primaries at various combinations of $\rho$ and $\Delta$mag. The color scale is logarithmic. The predictions are from Monte Carlo simulations that use  \citet{raghavan} field-binary statistics. The 20 control-star companions detected at 832~nm with separations under 1.2" are also shown, revealing the conspicuous absence of detected companions at $0.1"\lesssim\rho\lesssim1.2$" and $\Delta$mag~$\lesssim3$ (the region bounded with a thick dashed white line). The simulations predict no such deficit, which we show to be the result of a selection bias that prevented the observation of binaries in this parameter space. 
                }
                \label{fig:selection_bias}
            \end{figure*}

             The \citet{raghavan} statistics predict that we should have detected $23\pm3$ companions at 832~nm and $18\pm3$ at 562~nm around the FGK dwarfs in our sample. For comparison, we found that only 15 FGK control stars had companions within 1.2" at 832~nm (and just 8 at 562~nm), so there is a significant deficit of companions in our observations. This deficit cannot be explained by uncertainty in the contrast curves used in the simulations. As we discussed earlier, the contrast curves within $\rho < 0.1"$ underestimate our ability to detect close-in companions in good seeing, and since our simulations predict that there should be many companions within $\rho < 0.1"$, the totals predicted by our simulations are a lower limit.

            Fig.~\ref{fig:selection_bias} shows a two-dimensional histogram of the expected detection rate of synthetic companions as a function of both $\rho$ and $\Delta$mag, and to facilitate a direct comparison against the observations, it also plots $\rho$ and $\Delta$mag for all companions detected at 832~nm. This comparison reveals a striking disagreement between the simulations and observations. In particular, not a single observed companion to an FGK star has $0.1"\leq\rho\leq1.2"$ and $\Delta$mag$<3$, even though the simulations predict this to be a fertile region for finding companions. Indeed, our simulations predicted an average of $9\pm2$~companions with $0.1"\leq\rho\leq1.2"$ and $\Delta$mag$<3$; the minimum number of predicted companions in this parameter space was 4. 
            
            The absence of observed companions with $0.1"\leq\rho\leq1.2"$ and $\Delta$mag$<3$ for FGK stars is a stark deviation from the properties of \citet{raghavan} field binaries, and taken at face value, it would conflict with a line of studies of stellar multiplicity surveys of transiting-exoplanet host stars. However, we show in Sec.~\ref{sec:bias} that this surprising deficit is attributable to a selection bias in the creation of the control sample.

        \subsection{Origin of the bias} \label{sec:bias}

        The control stars were selected based on their stellar parameters in the TIC (specifically, their effective temperatures, radii, and distances), as well as the uncertainties of these parameters. However, before stellar parameters are calculated in the TIC workflow, two preliminary tests of the quality of the Gaia DR2 astrometry and photometry are performed; if a source passes both tests, it is assigned a Gaia quality flag ({\tt gaiaqflag}) of 1; if the source fails one or both tests, then {\tt gaiaqflag} = 0. 
    
        Although verifying the caliber of the astrometry and photometry might sound innocuous and even desirable, this process has deleterious consequences for close binaries. Previous studies \citep[e.g.,][]{deacon20, ziegler20} have already established that Gaia astrometry of close, unresolved binaries tends to have large residuals, quantified by the renormalized unit weight error statistic (RUWE). Indeed, the RUWE can be well over an order of magnitude larger for binaries with separations of $0.1" \lesssim \rho \lesssim 1"$ and magnitude differences $\lesssim2$~mag \citep[Fig.~4 in ][]{ziegler20}. As a result, quality cuts made on astrometry will penalize unresolved binaries in that parameter space, causing these systems to be disproportionately excluded in comparison to single stars, widely separated binaries, extremely close binaries ($\rho < 0.1"$), and binaries with very large $\Delta$~mag.
    
        To be clear, when a source fails either of the Gaia quality tests (i.e., {\tt gaiaqflag} = 0), it is still included in the TIC. But the stellar parameters of these sources tend to be incomplete and/or lack uncertainties compared to sources that pass the quality tests ({\tt gaiaqflag} = 1). This is because the TIC automatically computes stellar parameters only if {\tt gaiaqflag} = 1. While stars with {\tt gaiaqflag} = 0 might have stellar parameters and uncertainties, this happens only if they were obtained from other sources, such as spectroscopic catalogs or specially curated lists. Thus, by requiring the control stars to have valid $T_{eff}$, radii, and distances (along with uncertainties on each parameter), our selection criteria unintentionally excluded certain types of binaries from the control sample.

        \subsection{Testing the bias}

        \begin{figure*}
            \centering
            \includegraphics[width=\linewidth]{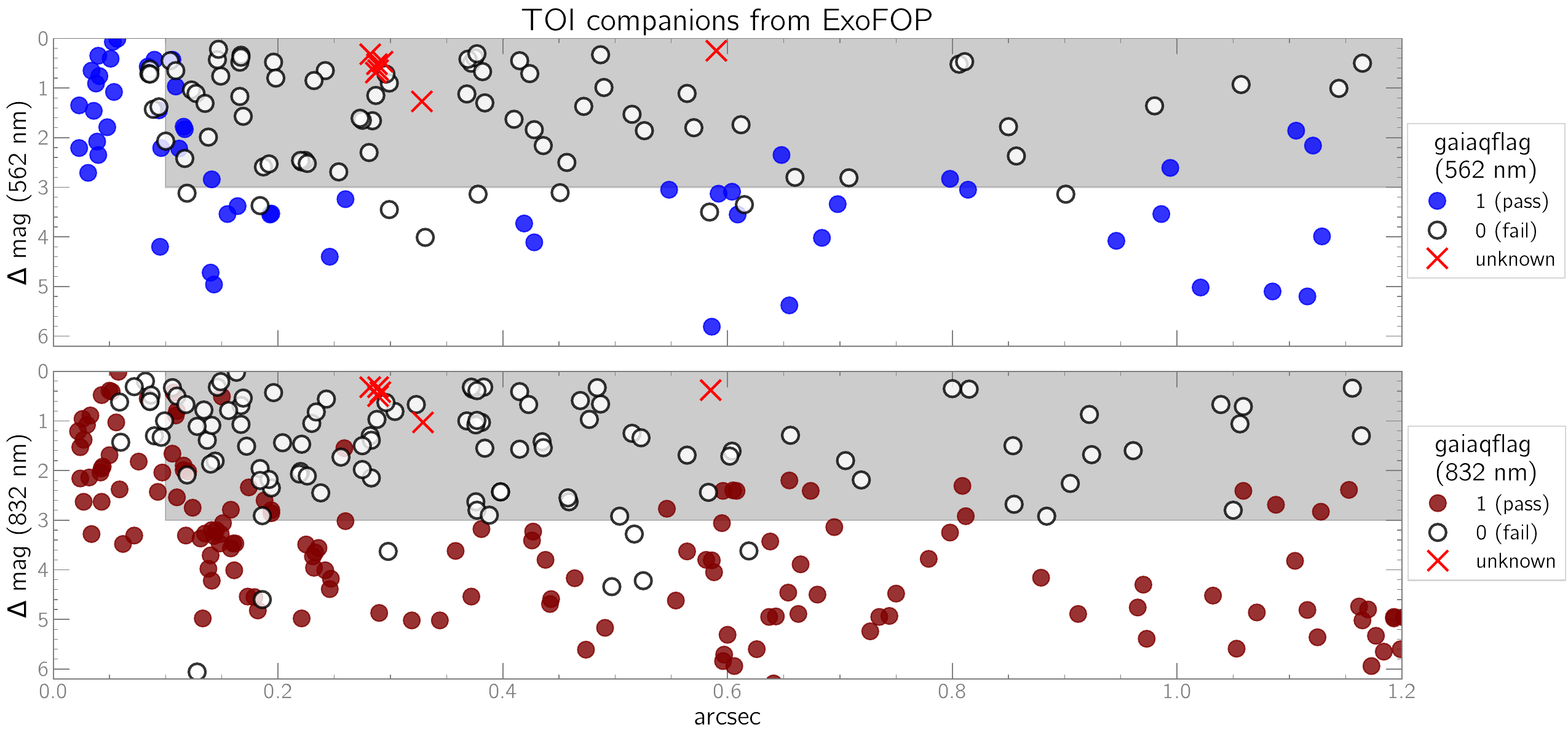}
            \caption{The dependence of {\tt gaiaqflag} on $\Delta$mag and $\rho$ in companion-hosting TOI objects. Binaries with $\Delta$mag $\lesssim3$ and $0.1" \lesssim \rho \lesssim 1.2"$ (represented with a shaded region in both panels) disproportionately fail the Gaia quality test. As we discuss in the text, sources that fail the Gaia quality test will lack the parameters used to assemble our control sample, creating a powerful selection bias against binaries within this parameter space. 
            \label{fig:exofop}}
        \end{figure*}

    \begin{figure}
        \centering
        \includegraphics[width=\columnwidth]{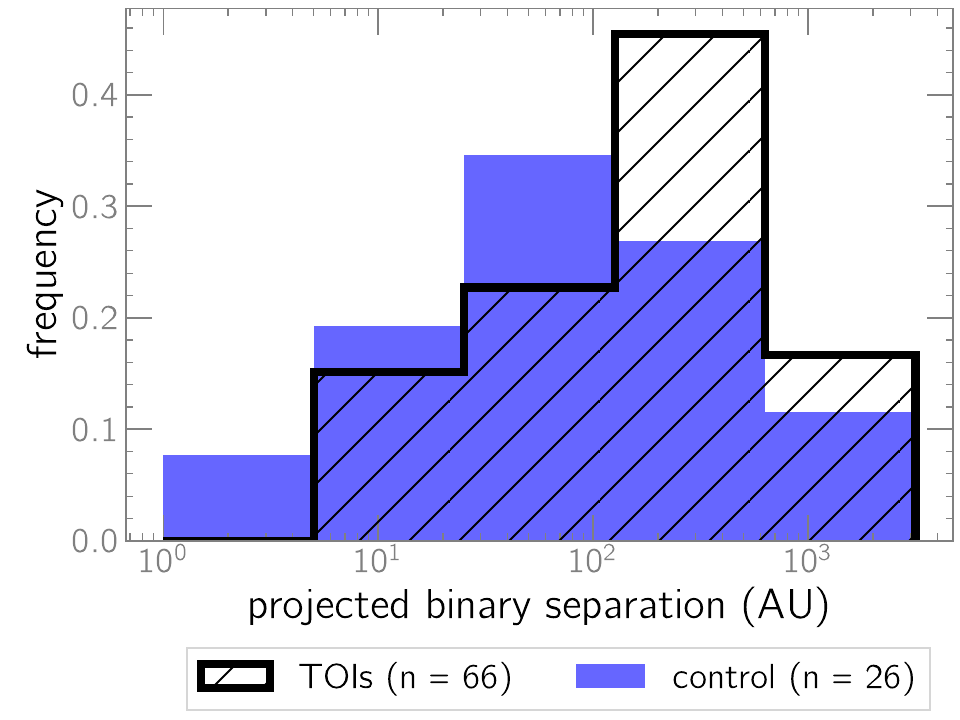}
        \includegraphics[width=\columnwidth]{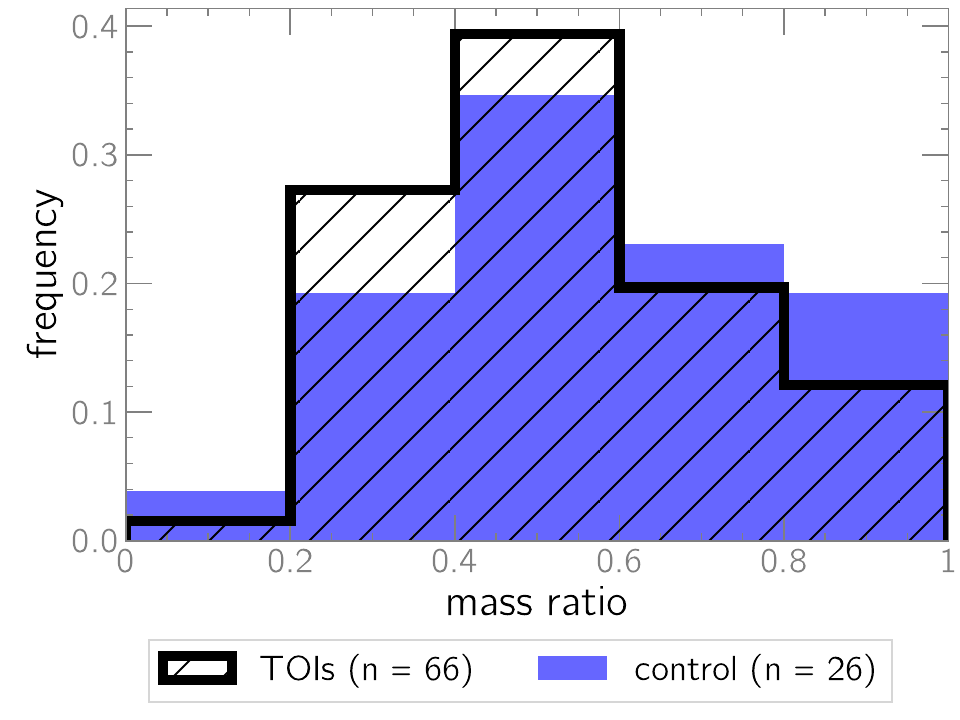}
        \caption{{\bf Top:} Projected binary separations for the detected companions in the control sample compared against those of the \citet{howell21} and \citet{lester21} TOI stars whose values of $\rho$ and $\Delta$mag place them outside the parameter space of the selection bias. A K-S test comparing these two distributions has a $p$-value of 0.07, indicating marginal statistical evidence of a difference. {\bf Bottom:} As with the top panel, except that mass ratios are compared. The corresponding K-S test has a $p$-value of 0.2, indicating no statistically significant difference. In both panels, we lift the restrictions on spectral type and $\rho$ used in the simulations, so all 26 detected companions around control stars are included.}
        \label{fig:separations_and_Q}
    \end{figure}
    
        As a test of this scenario, we downloaded from the ExoFOP database\footnote{\url{https://exofop.ipac.caltech.edu/tess/}} a list of speckle-detected companions to TOI stars at 562~nm and 832~nm \citep{https://doi.org/10.26134/exofop3}. Figure~\ref{fig:exofop} plots $\Delta$mag as a function of $\rho$ for these companions and uses different marker styles to illustrate the dependence of {\tt gaiaqflag} on both $\rho$ and $\Delta$mag. Consistent with \citet{deacon20} and \citet{ziegler20}, the stars with companions within $0.1" \lesssim \rho \lesssim 1"$ and $\Delta$~mag~(832~nm)$\lesssim2$ overwhelmingly show {\tt gaiaqflag} = 0. This result holds for companions detected at 562~nm, except that the sources with {\tt gaiaqflag} = 0 extend to $\Delta$mag~(562~nm)$\lesssim3$. 

        The next step in assessing the impact of this bias is to examine the impact of {\tt gaiaqflag} = 0 on our selection criteria. To that end, we inspected the TIC parameters of the companions shown in Figure~\ref{fig:exofop} to assess whether they differ depending on the value of {\tt gaiaqflag}. Almost every source with {\tt gaiaqflag} = 1 had valid parameters and uncertainties for the stellar radius, effective temperature, and distance. However, a very different picture emerges for the sources with {\tt gaiaqflag} = 0, of which only 42\% had estimates for all three of those parameters. Moreover, \textit{none} of the {\tt gaiaqflag} = 0 sources had valid uncertainties for all three of those parameters. The combination of incomplete TIC parameters and missing uncertainties means that it would be difficult or even impossible to select these stars for observation based solely on the criteria used to assemble the control sample.

        \subsection{Impact of the bias}
    
        The selection bias described above is insidiously subtle because it is caused by the combination of the TIC's quality-control tests and our relatively stringent selection criteria. In particular, it would normally seem reasonable to require, as we did, that sources have uncertainties on their stellar parameters, and we expect that this selection bias will afflict studies that select their targets based solely on their stellar parameters and their associated uncertainties in the TIC. With this in mind, it is worth considering how this bias could distort the scientific interpretation of the data. 
        
        Left uncorrected, this selection bias might lead to the unwarranted conclusion that the FGK control stars have a much lower companion fraction than the canonical value of $46\pm2\%$, based on the simple fact that simulations using \citet{raghavan} statistics significantly overpredict the number of observed companions. However, this deficit disappears once the simulations are modified to simulate the selection bias. After adjusting our simulations to disallow the creation of binaries with $0.1"\leq\rho\leq 1.2"$ and $\Delta$mag $< 3$~mag, we would expect to find, within 1.2", $14.7\pm2.3$ companions at 832~nm and $10.7\pm2.0$ at 562~nm, assuming a 46$\pm$2\% multiplicity rate from \citet{raghavan}. These predictions are in remarkably good agreement with the observed totals of 15 and 8 for those respective bandpasses. In short, the observed companion rate of the FGK control stars within 1.2" and $\Delta$mag $< 3$ mag agrees with the \citet{raghavan} statistics.
        
        Likewise, the FGK dwarfs in the control sample show an apparent enhancement of binaries with q$\sim$0.5 (Figure~\ref{fig:separations_and_Q}), a feature that is inconsistent with the \citet{raghavan} mass-ratio distribution. But here again, the selection bias is responsible for this apparent feature. When both components of a binary are on the main sequence and have similar masses, they will have similar luminosities and hence a small $\Delta$mag, making them vulnerable to this selection bias if their angular separations are wider than 0.1". Conversely, unequal-mass binaries will have larger $\Delta$mag, causing them to be overrepresented in our control sample.

       \subsection{Properties of detected binaries} \label{sec:companion_properties}

       Due to the potency of this selection bias (Figure~\ref{fig:selection_bias} and Figure~\ref{fig:exofop}), we can perform only a limited comparison between the control-sample binaries and companions to TOIs. We took TOI binaries from \citet{howell21} and \citet{lester21} and excluded those with $0.1"\leq\rho\leq 1.2"$ and $\Delta$mag $< 3$~mag in order to mimic the selection bias that afflicts the control sample. The upper panel in Figure~\ref{fig:separations_and_Q} presents a histogram of the projected separations of the FGK stars in the control sample, compared to the TOI binaries from \citet{howell21} and \citet{lester21}, while the lower panel compares their mass ratios. In both comparisons, we used a Kolomogorov-Smirnoff (K-S) test to evaluate how different the two distributions are from each other. The $p$-value of the K-S test indicates the statistical significance of the difference in the compared distributions, with largest significance for the smallest values of $p$ ($p=0.00$--$0.01$). The $p$-value for the mass-ratio comparison is 0.2, suggesting no significant difference between the control-star and TOI mass ratios subject to the aforementioned constraints. However, there is modest evidence ($p$ = 0.07) for a difference in the projected separations of the control stars and binary TOIs. Figure~\ref{fig:separations_and_Q} shows that the TOI companions have wider separations than the control-star companions; if true, this would be consistent with earlier findings \citep[e.g.][]{howell21, lester21} that exoplanet-hosting TOI binaries have wider separations than field binaries.

    \subsection{Statistical significance}
   
        Our sample size (n = 207) is relatively small compared to the $>5000$ TOIs at the time of writing, and the selection bias described earlier complicates comparisons between the control sample and the corresponding TOIs. For example, if the FGK control stars' multiplicity rate or binary separations varied dramatically from the \citet{raghavan} statistics, would we be able to tell using the available data?

        We used our simulations to explore this issue in several ways. First, we tried predicting the number of detectable companions using the orbital period distribution from \citet{lester21}. That study found that companions to TOI host stars had wider binary separations than field stars and reported a log-normal orbital-period distribution with $\mu = 6.2$~d and $\sigma=1.2$. We reran our simulations using this orbital-period distribution while disallowing the creation of binaries that would be subject to the selection bias. The \citet{lester21} distribution predicted $19.3\pm2.6$ red and $12.1\pm2.3$ blue companions to FGK stars within 1.2". For reference, the observed totals were 15 and 8, respectively, so the \citet{lester21} period distribution overpredicts the number of companions in both filters.

        Although 18 M-dwarfs were included in our control sample, we find this number to be too small for useful statistical inference. Three of those stars (all M1.5 or earlier) harbor companions. Following \citet{clark22}, we adjusted our simulations to use the \citet{dk13} orbital-period distribution for M-dwarf companions, a flat mass-ratio distribution, and a 27\% multiplicity fraction. Using these statistics, our simulations predicted $1.4\pm0.5$ red companions and $1.1\pm0.5$ blue companions, accounting for the selection bias in our study. At face value, this would seem to be somewhat at odds with the lack of close-in companions identified by \citet{clark22} in their high-resolution imaging of M-dwarf TOIs.

        However, it is worth remembering that we estimate spectral types by comparing the  effective temperatures reported in the TIC with the \citet{pm13} sequence. The TIC's effective temperatures have uncertainties that propagate to the inferred spectral type, and in the case of early M stars, this means that we cannot rule out the possibility that they are late K stars that have been erroneously excluded from our analysis of the FGK control stars. For example, for one of our control stars (TIC 296781193), $T_{eff} = 3882\pm157$~K, and while the inferred spectral type from \citet{pm13} is M0, the uncertainty means that the spectral type could be as early as K8 or as late as M0.5. The other two M-dwarfs with companions are similarly close to the threshold between spectral types M and K. Thus, the nominal tally of three companions to M-dwarfs in the control sample is rather uncertain.

    \section{Conclusion}

        We obtained speckle observations of 207 stars selected because their effective temperature, radius, and distance in the TIC closely resemble the parameters of a corresponding TOI. Our goal was to create a control sample for TOIs in order to assess whether the occurrence rate and properties of stellar companions to TOIs differ from otherwise identical stars that are not known (or suspected) to host short-period transiting exoplanets.
        
        We identified a significant selection bias against the inclusion of certain types of close binaries during the creation of the control sample. The bias occurs because the presence of an unresolved companion causes a source in the TIC to fail preliminary quality tests if the binary separation is in the range $0.1" \lesssim \rho \lesssim 1.2"$ and $\Delta$mag $\lesssim 3$~mag. Sources that fail the TIC quality test tend to have incomplete stellar parameters and often lack uncertainties on the available stellar parameters, so normally uncontroversial selection criteria can have the unintended effect of systematically excluding these close, unresolved binaries. Since this bias is the result of target-selection criteria that rely solely upon TIC stellar parameters and uncertainties, it can be mitigated by using selection criteria based on other sources, and this would be a worthy objective for an improved control sample.

        After accounting for this selection bias, our simulations using \citet{raghavan} binary statistics for field FGK stars correctly predicted the number of companions that we detected around our control stars. Moreover, we found that the mass-ratio distribution of the control-star companions agrees with that of binary TOI host stars whose $\rho$ and $\Delta$mag place them outside the parameter space of the selection bias. After correction for the bias introducted into the control sample, we still find modest evidence that the physical separation of binary stars in TOIs is wider than the physical separation of binary components in transiting planet host stars.

    \begin{acknowledgements}
        We thank the anonymous referee for an insightful review that led to the improvement of this manuscript. 

    \end{acknowledgements}

\bibliography{bib.bib}

\startlongtable
\begin{deluxetable*}{ccc|cc|cc|cc}
\label{table:control_sample}
\tablecaption{Correspondence between TOIs and control stars}
\tablehead{
    \colhead{TIC ID} & 
    \colhead{TOI} & 
    \colhead{Binary} & 
    \multicolumn{2}{c}{Distance (pc)} &
    \multicolumn{2}{c}{Radius (R$_{\odot}$)} &
    \multicolumn{2}{c}{$T_{eff}$ (K)} \\
    \colhead{(control star)} &
    \colhead{} &
    \colhead{(control star)} &
    \colhead{control} &
    \colhead{TOI} &
    \colhead{control} &
    \colhead{TOI} &
    \colhead{control} &
    \colhead{TOI}
}
\startdata
194461161 & 280 & B & 99.95$\pm$0.95 & 99.66$\pm$0.30 & 0.787$\pm$0.037 & 0.783$\pm$0.039 & 5410$\pm$110 & 5450$\pm$190 \\
301482610 & 297 & B & 455.3$\pm$5.2 & 461.5$\pm$5.3 & 1.123$\pm$0.071 &  1.09  $\pm$  0  & 4960$\pm$120 & 4870$\pm$120 \\
196383895 & 518 & B & 161.0$\pm$1.6 & 159.8$\pm$1.2 & 1.132$\pm$0.059 & 1.14$\pm$0.15 & 5610$\pm$130 & 5600$\pm$190 \\
377058463 & 586 & B & 256.6$\pm$4.1 & 253.0$\pm$2.5 & 2.387$\pm$0.070 & 2.420$\pm$0.070 & 9510$\pm$150 & 9510$\pm$180 \\
264485499 & 692 & B & 483$\pm$22 & 482$\pm$19 & 2.68$\pm$0.15 & 2.74$\pm$0.13 & 9480$\pm$170 & 9620$\pm$140 \\
376688975 & 704 & B & 29.69$\pm$0.13 & 29.842$\pm$0.024 & 0.510$\pm$0.015 & 0.504$\pm$0.015 & 3690$\pm$160 & 3625$\pm$69 \\
355691670 & 772 & B & 135.55$\pm$0.74 & 135.37$\pm$0.96 & 0.814$\pm$0.040 & 0.820$\pm$0.050 & 5170$\pm$100 & 5180$\pm$180 \\
296781193 & 833 & B & 41.31$\pm$0.24 & 41.715$\pm$0.042 & 0.617$\pm$0.018 & 0.603$\pm$0.018 & 3880$\pm$160 & 3965$\pm$72 \\
301482228 & 1493 & B & 373$\pm$22 & 371.4$\pm$6.2 & 2.88$\pm$0.21 & 2.90$\pm$0.14 & 6030$\pm$100 & 6060$\pm$130 \\
354442089 & 1669 & B & 110.90$\pm$0.76 & 111.28$\pm$0.37 & 1.080$\pm$0.066 & 1.070$\pm$0.050 & 5540$\pm$140 & 5550$\pm$130 \\
411551642 & 1805 & B & 142.76$\pm$0.81 & 142.85$\pm$0.43 & 0.843$\pm$0.046 & 0.860$\pm$0.040 & 5260$\pm$120 & 5230$\pm$110 \\
85274754 & 1966 & B & 255.1$\pm$6.6 & 254.2$\pm$3.2 & 2.29$\pm$0.10 & 2.31$\pm$0.10 & 6430$\pm$100 & 6390$\pm$260 \\
372086900 & 2049 & B & 547.9$\pm$8.4 & 571.5$\pm$9.0 & 4.78$\pm$0.22 & 3.48$\pm$0.21 & 6670$\pm$180 & 6050$\pm$150 \\
83958546 & 2101 & B & 178.1$\pm$2.3 & 178.43$\pm$0.59 & 0.998$\pm$0.046 & 1.000$\pm$0.050 & 5640$\pm$100 & 5620$\pm$130 \\
269390255 & 2166 & B & 346.3$\pm$9.3 & 348.6$\pm$3.5 & 1.98$\pm$0.10 & 1.990$\pm$0.090 & 6570$\pm$170 & 6520$\pm$120 \\
716026635 & 2193 & B & 338.2$\pm$7.2 & 337.1$\pm$2.4 & 1.205$\pm$0.057 & 1.197$\pm$0.055 & 6190$\pm$150 & 6080$\pm$130 \\
2100594 & 2196 & B & 260.3$\pm$3.6 & 259.9$\pm$2.0 & 1.028$\pm$0.051 & 1.029$\pm$0.050 & 5720$\pm$160 & 5730$\pm$130 \\
91277756 & 2458 & B & 112.1$\pm$2.3 & 112.96$\pm$0.54 & 1.333$\pm$0.066 & 1.330$\pm$0.060 & 6080$\pm$150 & 6090$\pm$120 \\
306125356 & 2523 & B & 166.0$\pm$1.4 & 166.3$\pm$1.0 & 1.115$\pm$0.060 & 1.120$\pm$0.050 & 5790$\pm$130 & 5810$\pm$270 \\
267686220 & 3159 & B & 427$\pm$27 & 428.1$\pm$5.3 & 1.82$\pm$0.15 & 1.810$\pm$0.090 & 5700$\pm$120 & 5730$\pm$130 \\
91597865 & 3415 & B & 1048$\pm$67 & 1053$\pm$38 & 2.54$\pm$0.19 & 2.56$\pm$0.15 & 5940$\pm$140 & 5940$\pm$120 \\
14899687 & 3755 & B & 320.5$\pm$5.3 & 322.4$\pm$3.0 & 1.074$\pm$0.055 & 1.080$\pm$0.060 & 5350$\pm$110 & 5350$\pm$320 \\
241257501 & 4244 & B & 770$\pm$77 & 768$\pm$21 & 1.61$\pm$0.18 & 1.61$\pm$0.10 & 6300$\pm$130 & 6310$\pm$200 \\
437327600 & 4305 & B & 159.2$\pm$2.3 & 159.01$\pm$0.82 & 2.79$\pm$0.13 & 2.75$\pm$0.12 & 6160$\pm$140 & 6200$\pm$130 \\
140343515 & 4441 & B & 628$\pm$18 & 627$\pm$11 & 1.990$\pm$0.087 & 1.998$\pm$0.083 & 7340$\pm$220 & 7310$\pm$130 \\
96876685 & 5528 & B & 103.82$\pm$0.42 & 101.63$\pm$0.37 & 0.546$\pm$0.016 & 0.515$\pm$0.015 & 3610$\pm$160 & 3690$\pm$160 \\
416670433 & 173 &    & 151.8$\pm$1.2 & 152.02$\pm$0.64 & 1.465$\pm$0.074 & 1.470$\pm$0.060 & 6440$\pm$470 & 6410$\pm$200 \\
247873150 & 201 &    & 113.79$\pm$0.70 & 113.83$\pm$0.35 & 1.308$\pm$0.056 & 1.330$\pm$0.060 & 6460$\pm$160 & 6462$\pm$83 \\
437761325 & 209 &    & 113.27$\pm$0.66 & 113.39$\pm$0.27 & 0.742$\pm$0.051 & 0.743$\pm$0.050 & 4900$\pm$130 & 4890$\pm$180 \\
101831762 & 214 &    & 39.120$\pm$0.043 & 38.960$\pm$0.038 & 0.783$\pm$0.044 & 0.798$\pm$0.050 & 5260$\pm$120 & 5350$\pm$140 \\
23972719 & 229 &    & 211.3$\pm$1.4 & 211.2$\pm$1.8 & 0.885$\pm$0.057 & 0.875$\pm$0.057 & 5180$\pm$130 &  5100  $\pm$  0  \\
238431974 & 248 &    & 75.44$\pm$0.24 & 75.95$\pm$0.14 & 1.101$\pm$0.052 & 1.121$\pm$0.052 & 5730$\pm$120 & 5712$\pm$57 \\
65577518 & 262 &    & 43.572$\pm$0.076 & 43.93$\pm$0.12 & 0.832$\pm$0.037 & 0.849$\pm$0.052 & 5241$\pm$98 & 5303$\pm$21 \\
9006549 & 296 &    & 465.9$\pm$5.9 & 471.8$\pm$4.8 & 0.956$\pm$0.054 &  0.95  $\pm$  0  & 5400$\pm$120 & 5290$\pm$120 \\
230665144 & 328 &    & 659.6$\pm$8.2 & 660$\pm$14 & 1.205$\pm$0.057 & 1.200$\pm$0.060 & 6080$\pm$140 & 6070$\pm$130 \\
462606719 & 333 &    & 351.3$\pm$3.5 & 352.3$\pm$4.7 & 1.130$\pm$0.059 & 1.130$\pm$0.050 & 6150$\pm$140 & 6150$\pm$130 \\
414828178 & 458 &    & 77.20$\pm$0.31 & 76.70$\pm$0.21 & 0.821$\pm$0.047 & 0.830$\pm$0.053 & 5200$\pm$120 & 5160$\pm$180 \\
102205698 & 561 &    & 85.80$\pm$0.34 & 85.80$\pm$0.50 & 0.829$\pm$0.046 & 0.840$\pm$0.050 & 5320$\pm$120 & 5390$\pm$190 \\
406755887 & 569 &    & 154.9$\pm$1.2 & 156.23$\pm$0.75 & 1.556$\pm$0.063 & 1.470$\pm$0.070 & 6920$\pm$120 & 6850$\pm$200 \\
458447712 & 618 &    & 235.4$\pm$2.1 & 234.1$\pm$1.9 & 1.421$\pm$0.077 & 1.430$\pm$0.070 & 6510$\pm$160 & 6520$\pm$130 \\
96955726 & 626 &    & 441.7$\pm$8.4 & 437.2$\pm$8.4 & 2.090$\pm$0.072 & 2.320$\pm$0.080 & 8180$\pm$240 & 8490$\pm$180 \\
233655269 & 644 &    & 204.27$\pm$0.99 & 203.5$\pm$1.1 & 1.249$\pm$0.062 & 1.25$\pm$0.27 & 5940$\pm$130 & 5950$\pm$660 \\
4200768 & 695 &    & 144.74$\pm$0.66 & 144.05$\pm$0.54 & 0.804$\pm$0.041 & 0.810$\pm$0.040 & 5560$\pm$170 & 5590$\pm$190 \\
301797581 & 724 &    & 95.65$\pm$0.38 & 95.35$\pm$0.25 & 0.873$\pm$0.057 & 0.880$\pm$0.050 & 5350$\pm$150 & 5330$\pm$130 \\
357168810 & 763 &    & 96.11$\pm$0.36 & 95.13$\pm$0.45 & 0.910$\pm$0.069 & 0.907$\pm$0.047 & 5820$\pm$200 & 5770$\pm$190 \\
331609888 & 776 &    & 27.691$\pm$0.029 & 27.170$\pm$0.032 & 0.548$\pm$0.016 & 0.533$\pm$0.016 & 3680$\pm$160 & 3806$\pm$66 \\
25304963 & 789 &    & 43.241$\pm$0.075 & 43.407$\pm$0.058 & 0.371$\pm$0.011 & 0.371$\pm$0.011 & 3450$\pm$160 & 3471$\pm$64 \\
337219186 & 806 &    & 104.41$\pm$0.43 & 104.44$\pm$0.22 & 0.624$\pm$0.057 & 0.632$\pm$0.058 & 4140$\pm$180 & 4130$\pm$170 \\
10996450 & 815 &    & 59.90$\pm$0.17 & 59.71$\pm$0.13 & 0.754$\pm$0.053 & 0.760$\pm$0.040 & 4870$\pm$130 & 4950$\pm$110 \\
80884729 & 830 &    & 222.6$\pm$1.8 & 223.0$\pm$1.1 & 1.041$\pm$0.050 & 1.043$\pm$0.050 & 5900$\pm$130 & 5900$\pm$190 \\
27197662 & 851 &    & 153.9$\pm$1.1 & 154.5$\pm$2.3 & 0.776$\pm$0.061 & 0.770$\pm$0.040 & 5460$\pm$140 & 5490$\pm$140 \\
306067089 & 908 &    & 177.3$\pm$1.3 & 175.75$\pm$0.64 & 1.045$\pm$0.055 & 1.050$\pm$0.060 & 5540$\pm$120 & 5500$\pm$76 \\
412053583 & 1001 &    & 295.9$\pm$4.5 & 295.9$\pm$5.9 & 2.015$\pm$0.088 & 2.010$\pm$0.090 & 7070$\pm$420 & 7070$\pm$130 \\
94366803 & 1136 &    & 84.34$\pm$0.30 & 84.54$\pm$0.16 & 0.971$\pm$0.046 & 0.980$\pm$0.050 & 5720$\pm$120 & 5730$\pm$130 \\
194465450 & 1187 &    & 221.5$\pm$3.9 & 221.1$\pm$5.0 & 1.406$\pm$0.072 & 1.400$\pm$0.070 & 6410$\pm$140 & 6390$\pm$120 \\
311384795 & 1208 &    & 135.16$\pm$0.91 & 134.76$\pm$0.36 & 0.814$\pm$0.043 & 0.823$\pm$0.041 & 5660$\pm$150 & 5630$\pm$190 \\
345670001 & 1219 &    & 299.3$\pm$4.5 & 301.3$\pm$2.2 & 1.594$\pm$0.068 & 1.590$\pm$0.060 & 6730$\pm$150 & 6730$\pm$200 \\
326581975 & 1221 &    & 138.32$\pm$0.55 & 138.41$\pm$0.39 & 1.020$\pm$0.050 & 1.018$\pm$0.049 & 5830$\pm$130 & 5800$\pm$190 \\
139477575 & 1224 &    & 36.922$\pm$0.063 & 37.331$\pm$0.060 & 0.410$\pm$0.012 & 0.404$\pm$0.012 & 3460$\pm$160 & 3442$\pm$64 \\
186990840 & 1247 &    & 74.19$\pm$0.32 & 73.87$\pm$0.14 & 1.082$\pm$0.052 & 1.080$\pm$0.050 & 5730$\pm$120 & 5710$\pm$110 \\
457164920 & 1274 &    & 178.6$\pm$1.1 & 178.14$\pm$0.96 & 0.797$\pm$0.053 & 0.800$\pm$0.060 & 4940$\pm$140 & 4970$\pm$170 \\
298246831 & 1333 &    & 200.29$\pm$0.99 & 200.5$\pm$1.2 & 1.796$\pm$0.097 & 1.800$\pm$0.080 & 6640$\pm$170 & 6700$\pm$240 \\
22499582 & 1413 &    & 115.90$\pm$0.90 & 115.4$\pm$1.1 & 0.891$\pm$0.059 & 0.890$\pm$0.050 & 5310$\pm$140 & 5430$\pm$140 \\
115619561 & 1448 &    & 74.08$\pm$0.33 & 73.33$\pm$0.17 & 0.382$\pm$0.011 & 0.382$\pm$0.011 & 3390$\pm$160 & 3390$\pm$160 \\
10960327 & 1451 &    & 92.43$\pm$0.35 & 91.94$\pm$0.22 & 1.021$\pm$0.056 & 1.030$\pm$0.050 & 5810$\pm$140 & 5780$\pm$130 \\
417115924 & 1453 &    & 78.65$\pm$0.23 & 78.74$\pm$0.12 & 0.712$\pm$0.052 & 0.710$\pm$0.050 & 4960$\pm$140 & 4920$\pm$130 \\
1948923 & 1481 &    & 1031$\pm$61 & 1026$\pm$56 & 4.05$\pm$0.27 & 4.02$\pm$0.24 & 8460$\pm$180 & 8330$\pm$130 \\
453223414 & 1491 &    & 206.0$\pm$1.5 & 204.9$\pm$1.6 & 1.226$\pm$0.058 & 1.230$\pm$0.060 & 5860$\pm$200 & 5900$\pm$140 \\
36956523 & 1494 &    & 278.0$\pm$5.2 & 278.5$\pm$3.3 & 1.478$\pm$0.069 & 1.480$\pm$0.070 & 6420$\pm$120 & 6430$\pm$130 \\
351368610 & 1546 &    & 210.4$\pm$1.8 & 210.1$\pm$5.8 & 1.407$\pm$0.065 & 1.400$\pm$0.080 & 6210$\pm$130 & 6220$\pm$110 \\
69750913 & 1548 &    & 96.26$\pm$0.67 & 96.52$\pm$0.44 & 1.205$\pm$0.055 & 1.210$\pm$0.070 & 5820$\pm$120 & 5860$\pm$140 \\
411888197 & 1563 &    & 51.17$\pm$0.13 & 51.04$\pm$0.11 & 0.743$\pm$0.057 & 0.740$\pm$0.060 & 4570$\pm$120 & 4580$\pm$130 \\
224305482 & 1605 &    & 158.58$\pm$0.66 & 159.7$\pm$1.5 & 1.559$\pm$0.077 & 1.590$\pm$0.080 & 5600$\pm$130 & 5600$\pm$150 \\
1521852 & 1620 &    & 207.1$\pm$1.9 & 207.5$\pm$1.0 & 1.509$\pm$0.064 & 1.490$\pm$0.070 & 6630$\pm$130 & 6590$\pm$140 \\
256779898 & 1655 &    & 159.8$\pm$1.0 & 165.31$\pm$0.99 & 1.151$\pm$0.064 & 1.150$\pm$0.060 & 5550$\pm$100 & 5410$\pm$150 \\
385555166 & 1677 &    & 197.6$\pm$2.3 & 197.1$\pm$3.0 & 1.492$\pm$0.068 & 1.490$\pm$0.090 & 5880$\pm$110 & 5920$\pm$110 \\
136721562 & 1685 &    & 37.944$\pm$0.041 & 37.615$\pm$0.073 & 0.460$\pm$0.013 & 0.462$\pm$0.014 & 3430$\pm$160 & 3460$\pm$160 \\
348663808 & 1693 &    & 30.874$\pm$0.059 & 30.795$\pm$0.039 & 0.459$\pm$0.014 & 0.461$\pm$0.014 & 3440$\pm$160 & 3470$\pm$160 \\
410698188 & 1699 &    & 205.4$\pm$1.5 & 205.3$\pm$1.9 & 1.040$\pm$0.052 & 1.040$\pm$0.060 & 5680$\pm$170 & 5650$\pm$130 \\
188586505 & 1716 &    & 104.45$\pm$0.53 & 104.16$\pm$0.81 & 1.218$\pm$0.057 & 1.240$\pm$0.070 & 5960$\pm$130 & 5880$\pm$140 \\
270167851 & 1732 &    & 74.13$\pm$0.26 & 74.76$\pm$0.23 & 0.626$\pm$0.019 & 0.630$\pm$0.020 & 3900$\pm$160 & 3880$\pm$160 \\
439915868 & 1743 &    & 41.58$\pm$0.13 & 41.276$\pm$0.089 & 0.3187$\pm$0.0096 & 0.320$\pm$0.010 & 3300$\pm$160 & 3280$\pm$160 \\
47284734 & 1750 &    & 162.1$\pm$1.1 & 162.46$\pm$0.84 & 0.767$\pm$0.062 & 0.770$\pm$0.050 & 4780$\pm$130 & 4770$\pm$120 \\
144383425 & 1752 &    & 102.24$\pm$0.39 & 103.02$\pm$0.34 & 0.528$\pm$0.016 & 0.530$\pm$0.020 & 3640$\pm$160 & 3650$\pm$160 \\
318887102 & 1754 &    & 81.36$\pm$0.31 & 81.55$\pm$0.15 & 0.589$\pm$0.018 & 0.590$\pm$0.020 & 3850$\pm$160 & 3850$\pm$160 \\
256836445 & 1835 &    & 32.349$\pm$0.063 & 32.159$\pm$0.057 & 0.770$\pm$0.046 & 0.782$\pm$0.038 & 5180$\pm$120 & 5300$\pm$110 \\
311659336 & 1846 &    & 47.76$\pm$0.13 & 47.250$\pm$0.096 & 0.406$\pm$0.012 & 0.410$\pm$0.010 & 3550$\pm$160 & 3510$\pm$160 \\
367851162 & 1849 &    & 189.9$\pm$1.6 & 189.1$\pm$3.1 & 1.133$\pm$0.058 & 1.140$\pm$0.060 & 5600$\pm$110 & 5600$\pm$130 \\
17930632 & 1858 &    & 206.5$\pm$1.8 & 205.4$\pm$1.0 & 0.766$\pm$0.063 & 0.770$\pm$0.070 & 4230$\pm$110 & 4220$\pm$120 \\
91928178 & 1860 &    & 46.14$\pm$0.14 & 45.864$\pm$0.065 & 0.908$\pm$0.043 & 0.930$\pm$0.039 & 5680$\pm$110 & 5670$\pm$100 \\
412154086 & 1963 &    & 157.5$\pm$1.5 & 158.1$\pm$2.4 & 1.180$\pm$0.065 & 1.190$\pm$0.060 & 5450$\pm$130 & 5390$\pm$120 \\
139463870 & 1994 &    & 516$\pm$18 & 515.6$\pm$9.0 & 2.22$\pm$0.13 & 2.223$\pm$0.074 & 7270$\pm$150 & 7370$\pm$220 \\
435882484 & 2015 &    & 47.55$\pm$0.15 & 47.34$\pm$0.11 & 0.3131$\pm$0.0093 & 0.320$\pm$0.010 & 3240$\pm$160 & 3220$\pm$160 \\
440633965 & 2019 &    & 197.8$\pm$2.0 & 198.01$\pm$0.82 & 1.797$\pm$0.091 & 1.798$\pm$0.090 & 5650$\pm$130 & 5590$\pm$120 \\
467785978 & 2035 &    & 361.6$\pm$9.1 &  359.404  $\pm$  0  & 2.325$\pm$0.084 &  2.35  $\pm$  0  & 9240$\pm$140 &  9387  $\pm$  0  \\
117796162 & 2050 &    & 113.82$\pm$0.42 & 113.73$\pm$0.39 & 0.857$\pm$0.039 & 0.880$\pm$0.040 & 5990$\pm$140 & 5990$\pm$160 \\
102210132 & 2074 &    & 132.8$\pm$2.9 & 131.28$\pm$0.42 & 1.600$\pm$0.076 & 1.600$\pm$0.067 & 6560$\pm$120 & 6590$\pm$130 \\
27433678 & 2091 &    & 70.01$\pm$0.15 & 70.21$\pm$0.13 & 1.164$\pm$0.054 & 1.150$\pm$0.054 & 5690$\pm$110 & 5810$\pm$120 \\
430368654 & 2092 &    & 178.0$\pm$1.1 & 176.8$\pm$1.0 & 1.059$\pm$0.056 & 1.050$\pm$0.050 & 5440$\pm$130 & 5470$\pm$110 \\
286992966 & 2134 &    & 22.528$\pm$0.019 & 22.620$\pm$0.016 & 0.736$\pm$0.066 & 0.770$\pm$0.060 & 4600$\pm$140 & 4410$\pm$120 \\
158587692 & 2220 &    & 357.2$\pm$3.6 & 356.5$\pm$3.7 & 1.206$\pm$0.068 & 1.209$\pm$0.065 & 5640$\pm$140 & 5620$\pm$270 \\
444938803 & 2234 &    & 592$\pm$15 & 558$\pm$15 & 2.65$\pm$0.12 & 2.69$\pm$0.15 & 6140$\pm$110 & 6100$\pm$130 \\
436930863 & 2256 &    & 118.36$\pm$0.51 &  117.892  $\pm$  0  & 0.808$\pm$0.039 &  0.81  $\pm$  0  & 5310$\pm$100 &  5270  $\pm$  0  \\
23972248 & 2259 &    & 122.8$\pm$1.6 & 122.86$\pm$0.43 & 1.721$\pm$0.073 & 1.740$\pm$0.080 & 6640$\pm$120 & 6690$\pm$130 \\
24910668 & 2268 &    & 172.5$\pm$1.4 & 174.00$\pm$0.79 & 1.140$\pm$0.054 & 1.135$\pm$0.046 & 5900$\pm$130 & 5910$\pm$100 \\
97572659 & 2273 &    & 111.8$\pm$1.3 & 110.18$\pm$0.33 & 1.149$\pm$0.052 & 1.151$\pm$0.050 & 6430$\pm$140 & 6500$\pm$130 \\
283869896 & 2381 &    & 421.7$\pm$3.6 & 422.2$\pm$3.9 & 1.154$\pm$0.075 & 1.166$\pm$0.059 & 5330$\pm$140 & 5300$\pm$5300 \\
281914654 & 2420 &    & 442.7$\pm$8.4 & 435.2$\pm$7.2 & 2.11$\pm$0.11 & 2.20$\pm$0.12 & 5940$\pm$160 & 5710$\pm$130 \\
159088002 & 2497 &    & 285.7$\pm$3.3 & 285.3$\pm$3.5 & 2.50$\pm$0.10 & 2.50$\pm$0.11 & 6720$\pm$170 & 6720$\pm$170 \\
60960932 & 2501 &    & 94.21$\pm$0.50 & 93.88$\pm$0.18 & 0.660$\pm$0.062 & 0.659$\pm$0.067 & 4080$\pm$120 & 4070$\pm$130 \\
184692140 & 2541 &    & 747$\pm$37 & 738$\pm$16 & 3.53$\pm$0.20 & 3.52$\pm$0.12 & 7820$\pm$130 & 7840$\pm$140 \\
423725271 & 2614 &    & 315.6$\pm$3.5 &  315.5  $\pm$  0  & 1.016$\pm$0.052 &  1.0  $\pm$  0  & 5960$\pm$130 & 5900$\pm$6000 \\
81212238 & 2641 &    & 344.0$\pm$3.5 & 344.0$\pm$3.3 & 1.266$\pm$0.065 & 1.271$\pm$0.057 & 6300$\pm$150 & 6310$\pm$130 \\
141820840 & 2686 &    & 862$\pm$12 & 862$\pm$13 & 1.368$\pm$0.063 & 1.370$\pm$0.070 & 6260$\pm$120 & 6260$\pm$130 \\
22707360 & 2687 &    & 635.9$\pm$9.9 & 634$\pm$18 & 1.604$\pm$0.072 & 1.600$\pm$0.090 & 5660$\pm$100 & 5650$\pm$130 \\
117775653 & 2700 &    & 404.3$\pm$5.2 & 404.7$\pm$5.6 & 1.148$\pm$0.055 & 1.160$\pm$0.060 & 5630$\pm$120 & 5720$\pm$130 \\
331205514 & 2711 &    & 471.2$\pm$4.1 & 470.3$\pm$6.9 & 1.123$\pm$0.057 &  1.12  $\pm$  0  & 5550$\pm$180 & 5530$\pm$130 \\
249037563 & 2725 &    & 342.1$\pm$5.1 & 341.4$\pm$2.6 & 0.899$\pm$0.046 & 0.900$\pm$0.040 & 5920$\pm$130 & 5960$\pm$130 \\
139053927 & 2762 &    & 523$\pm$16 & 522.8$\pm$9.0 & 1.135$\pm$0.068 & 1.130$\pm$0.050 & 6270$\pm$140 & 6290$\pm$150 \\
5362070 & 2774 &    & 530.1$\pm$5.4 & 548$\pm$31 & 1.52$\pm$0.10 & 1.54$\pm$0.10 & 5000$\pm$130 & 4930$\pm$160 \\
354267216 & 2778 &    & 384.6$\pm$5.8 & 386.7$\pm$5.8 & 1.145$\pm$0.066 & 1.140$\pm$0.050 & 6330$\pm$590 & 6290$\pm$130 \\
80770158 & 2796 &    & 351$\pm$14 & 350.3$\pm$5.9 & 1.37$\pm$0.35 & 1.38$\pm$0.14 & 5470$\pm$110 & 5450$\pm$230 \\
16152210 & 2802 &    & 444.9$\pm$7.9 & 442$\pm$26 & 1.579$\pm$0.093 & 1.58$\pm$0.12 & 5680$\pm$140 & 5670$\pm$180 \\
417826992 & 2803 &    & 496.2$\pm$9.1 & 494.8$\pm$6.6 & 1.159$\pm$0.057 & 1.160$\pm$0.050 & 6400$\pm$130 & 6380$\pm$130 \\
285387244 & 2855 &    & 625$\pm$12 & 626.5$\pm$8.7 & 1.176$\pm$0.062 & 1.180$\pm$0.060 & 6060$\pm$130 & 6070$\pm$130 \\
236016211 & 2886 &    & 415.9$\pm$4.2 & 413.7$\pm$6.7 & 1.216$\pm$0.055 & 1.220$\pm$0.060 & 6150$\pm$130 & 6140$\pm$170 \\
345836977 & 2919 &    & 331.1$\pm$3.0 & 332.1$\pm$1.8 & 0.886$\pm$0.057 & 0.890$\pm$0.050 & 5460$\pm$140 & 5480$\pm$130 \\
151100143 & 2925 &    & 507$\pm$11 & 505$\pm$14 & 1.514$\pm$0.065 & 1.510$\pm$0.070 & 7180$\pm$180 & 7160$\pm$140 \\
411628261 & 2985 &    & 1263$\pm$48 & 1262$\pm$38 & 2.01$\pm$0.12 & 2.01$\pm$0.11 & 6420$\pm$120 & 6380$\pm$130 \\
80959830 & 2987 &    & 330.1$\pm$3.6 & 327.5$\pm$3.3 & 1.286$\pm$0.054 & 1.270$\pm$0.050 & 6980$\pm$140 & 7050$\pm$130 \\
68036178 & 2996 &    & 354.7$\pm$2.7 & 354.8$\pm$3.1 & 1.180$\pm$0.056 & 1.180$\pm$0.060 & 5940$\pm$120 & 5940$\pm$130 \\
4203392 & 3095 &    & 536.9$\pm$8.0 & 536.2$\pm$5.2 & 1.031$\pm$0.045 &  1.03  $\pm$  0  & 6020$\pm$100 & 6020$\pm$120 \\
116547873 & 3123 &    & 451.5$\pm$6.5 & 450.8$\pm$9.3 & 1.692$\pm$0.079 & 1.700$\pm$0.080 & 6370$\pm$140 & 6370$\pm$130 \\
266425483 & 3139 &    & 480$\pm$10 &  488.1  $\pm$ 0  & 1.45$\pm$0.10 &  1.43  $\pm$  0  & 5490$\pm$140 & 5450$\pm$460 \\
406474967 & 3177 &    & 335.3$\pm$2.6 & 335.2$\pm$2.1 & 0.893$\pm$0.053 & 0.890$\pm$0.050 & 5260$\pm$130 & 5250$\pm$120 \\
310240200 & 3196 &    & 321.2$\pm$3.5 & 321.3$\pm$2.0 & 0.907$\pm$0.050 & 0.910$\pm$0.050 & 5680$\pm$160 & 5670$\pm$130 \\
35301524 & 3202 &    & 374.7$\pm$5.7 & 374.0$\pm$5.6 & 1.025$\pm$0.052 & 1.020$\pm$0.050 & 5860$\pm$180 & 5830$\pm$130 \\
438462393 & 3231 &    & 1197$\pm$56 & 1206$\pm$37 & 3.26$\pm$0.19 & 3.28$\pm$0.15 & 7560$\pm$170 & 7590$\pm$140 \\
285177152 & 3267 &    & 716$\pm$15 & 714.8$\pm$9.3 & 1.410$\pm$0.073 & 1.410$\pm$0.080 & 5910$\pm$130 & 5890$\pm$130 \\
442965175 & 3294 &    & 287.0$\pm$2.8 &  286.4  $\pm$  0  & 1.349$\pm$0.084 & 1.390$\pm$0.050 & 5090$\pm$130 & 5003$\pm$95 \\
283226101 & 3305 &    & 395.9$\pm$3.4 & 394.4$\pm$3.5 & 0.865$\pm$0.049 & 0.860$\pm$0.060 & 5230$\pm$120 & 5300$\pm$140 \\
356735029 & 3342 &    & 340.8$\pm$2.8 & 340.8$\pm$4.3 & 1.092$\pm$0.054 & 1.090$\pm$0.050 & 5740$\pm$130 & 5750$\pm$130 \\
22499143 & 3346 &    & 138.05$\pm$0.80 & 132.57$\pm$0.75 & 0.698$\pm$0.059 & 0.700$\pm$0.060 & 4250$\pm$120 & 4320$\pm$120 \\
407965630 & 3350 &    & 112.22$\pm$0.53 & 112.09$\pm$0.34 & 0.916$\pm$0.092 & 0.900$\pm$0.060 & 4830$\pm$140 & 4790$\pm$120 \\
443756785 & 3364 &    & 277.9$\pm$3.1 & 277.9$\pm$2.5 & 1.450$\pm$0.070 & 1.450$\pm$0.070 & 5840$\pm$110 & 5810$\pm$130 \\
26654477 & 3370 &    & 549$\pm$10 & 549.1$\pm$8.5 & 1.466$\pm$0.066 & 1.470$\pm$0.060 & 6580$\pm$120 & 6590$\pm$130 \\
114924367 & 3406 &    & 894$\pm$20 & 890$\pm$23 & 1.693$\pm$0.083 & 1.690$\pm$0.080 & 6430$\pm$150 & 6440$\pm$130 \\
141203927 & 3516 &    & 378.9$\pm$3.4 & 378.3$\pm$3.4 & 1.595$\pm$0.077 & 1.580$\pm$0.080 & 5850$\pm$120 & 5820$\pm$130 \\
312232655 & 3589 &    & 572$\pm$14 & 574.1$\pm$9.0 & 1.605$\pm$0.062 & 1.610$\pm$0.080 & 7670$\pm$180 & 7660$\pm$150 \\
97715544 & 3635 &    & 1015$\pm$26 & 1013$\pm$19 & 2.15$\pm$0.12 & 2.16$\pm$0.11 & 5940$\pm$180 & 5970$\pm$120 \\
35116691 & 3638 &    & 424.1$\pm$3.9 & 425.3$\pm$4.8 & 1.151$\pm$0.065 & 1.150$\pm$0.060 & 5290$\pm$130 & 5260$\pm$120 \\
29943052 & 3653 &    & 622$\pm$10 & 622$\pm$12 & 1.122$\pm$0.055 &  1.12  $\pm$  0  & 5970$\pm$120 & 5980$\pm$120 \\
265988601 & 3686 &    & 255.1$\pm$2.2 & 255.8$\pm$2.1 & 1.258$\pm$0.071 & 1.240$\pm$0.070 & 5270$\pm$160 & 5260$\pm$160 \\
456211055 & 3699 &    & 771$\pm$20 & 771$\pm$16 & 1.681$\pm$0.081 & 1.660$\pm$0.070 & 7010$\pm$220 & 7050$\pm$220 \\
467791691 & 3738 &    & 716$\pm$19 & 716$\pm$33 & 1.759$\pm$0.085 & 1.76$\pm$0.13 & 6830$\pm$110 & 6810$\pm$140 \\
159027165 & 3757 &    & 181.32$\pm$0.97 & 181.1$\pm$1.2 & 0.640$\pm$0.019 & 0.640$\pm$0.020 & 3920$\pm$160 & 3900$\pm$160 \\
20924250 & 3766 &    & 449.7$\pm$5.4 & 451.7$\pm$5.0 & 0.995$\pm$0.060 &  0.99  $\pm$  0  & 5250$\pm$130 & 5240$\pm$120 \\
282353025 & 3807 &    & 420.1$\pm$8.4 & 421.7$\pm$4.6 & 1.479$\pm$0.073 & 1.470$\pm$0.070 & 5890$\pm$240 & 5910$\pm$130 \\
129984301 & 3817 &    & 1370$\pm$80 & 1387$\pm$33 & 2.11$\pm$0.15 &  2.06  $\pm$  0  & 6400$\pm$100 & 6310$\pm$120 \\
282263514 & 3823 &    & 347.4$\pm$5.0 & 353.4$\pm$8.0 & 1.115$\pm$0.052 & 1.100$\pm$0.050 & 6600$\pm$380 & 6390$\pm$110 \\
306120310 & 3859 &    & 783$\pm$19 & 785$\pm$33 & 1.478$\pm$0.087 & 1.480$\pm$0.090 & 6120$\pm$150 & 6120$\pm$110 \\
80425758 & 3891 &    & 331.2$\pm$4.0 & 332.0$\pm$2.3 & 0.937$\pm$0.045 & 0.940$\pm$0.050 & 5280$\pm$100 & 5270$\pm$110 \\
345050217 & 4000 &    & 879$\pm$12 & 878$\pm$16 & 1.927$\pm$0.071 & 1.900$\pm$0.070 & 7500$\pm$190 & 7490$\pm$220 \\
198279156 & 4029 &    & 273.7$\pm$2.3 & 274.4$\pm$3.3 & 1.189$\pm$0.060 & 1.180$\pm$0.050 & 5700$\pm$130 & 5710$\pm$110 \\
141264781 & 4081 &    & 435.3$\pm$6.0 & 439.9$\pm$6.5 & 2.57$\pm$0.13 & 2.54$\pm$0.13 & 5710$\pm$120 & 5710$\pm$150 \\
53090246 & 4090 &    & 249.6$\pm$1.5 & 250.98$\pm$0.81 & 0.819$\pm$0.057 & 0.820$\pm$0.060 & 4760$\pm$110 & 4740$\pm$120 \\
136951192 & 4095 &    & 560.4$\pm$5.8 & 559.6$\pm$4.4 & 1.091$\pm$0.053 & 1.088$\pm$0.048 & 5730$\pm$120 & 5740$\pm$110 \\
306346558 & 4104 &    & 1371$\pm$46 & 1367$\pm$51 & 2.20$\pm$0.12 & 2.20$\pm$0.13 & 6660$\pm$130 & 6680$\pm$130 \\
95957365 & 4116 &    & 502.3$\pm$9.2 &  504.9  $\pm$  0  & 1.643$\pm$0.083 &  1.63  $\pm$  0  & 5990$\pm$150 &  5980  $\pm$  0  \\
439899117 & 4153 &    & 425$\pm$10 & 423.8$\pm$3.6 & 1.680$\pm$0.094 & 1.670$\pm$0.080 & 6410$\pm$140 & 6410$\pm$230 \\
258194356 & 4156 &    & 213.1$\pm$1.2 & 213.7$\pm$1.0 & 1.035$\pm$0.049 & 1.040$\pm$0.050 & 6010$\pm$130 & 5990$\pm$130 \\
88252727 & 4166 &    & 257.7$\pm$2.2 & 258.0$\pm$1.5 & 0.961$\pm$0.047 & 0.960$\pm$0.050 & 5600$\pm$110 & 5600$\pm$140 \\
354297075 & 4183 &    & 135.47$\pm$0.76 & 135.1$\pm$1.0 & 1.245$\pm$0.052 & 1.245$\pm$0.055 & 6450$\pm$130 & 6410$\pm$200 \\
345641875 & 4190 &    & 120.87$\pm$0.79 & 121.49$\pm$0.45 & 0.819$\pm$0.051 & 0.817$\pm$0.051 & 5060$\pm$130 & 5070$\pm$130 \\
407571723 & 4192 &    & 507.2$\pm$8.2 & 507.6$\pm$7.0 & 1.441$\pm$0.071 & 1.440$\pm$0.070 & 6060$\pm$130 & 6060$\pm$130 \\
25392547 & 4231 &    & 1181$\pm$45 & 1176$\pm$40 & 1.675$\pm$0.081 & 1.670$\pm$0.070 & 8920$\pm$150 & 8950$\pm$290 \\
405248754 & 4248 &    & 653$\pm$15 & 654$\pm$12 & 2.10$\pm$0.11 & 2.09$\pm$0.10 & 5980$\pm$130 & 5936$\pm$71 \\
80434824 & 4262 &    & 338.4$\pm$8.8 & 337.2$\pm$3.5 & 1.293$\pm$0.066 & 1.290$\pm$0.060 & 6550$\pm$130 & 6610$\pm$130 \\
364904973 & 4267 &    & 965$\pm$16 & 964$\pm$17 & 1.479$\pm$0.069 &  1.48  $\pm$  0  & 6730$\pm$140 & 6720$\pm$120 \\
450297765 & 4278 &    & 689$\pm$14 & 688.8$\pm$9.9 & 1.396$\pm$0.087 &  1.39  $\pm$  0  & 5380$\pm$130 & 5480$\pm$120 \\
196790312 & 4285 &    & 747$\pm$13 & 746.4$\pm$8.9 & 1.234$\pm$0.059 & 1.230$\pm$0.050 & 6330$\pm$120 & 6330$\pm$130 \\
71625692 & 4290 &    & 227.3$\pm$2.3 & 228.13$\pm$0.94 & 0.799$\pm$0.064 & 0.800$\pm$0.040 & 5300$\pm$110 & 5310$\pm$130 \\
22571537 & 4296 &    & 115.63$\pm$0.67 & 115.72$\pm$0.51 & 0.896$\pm$0.052 & 0.899$\pm$0.057 & 4990$\pm$110 & 4990$\pm$120 \\
159033047 & 4309 &    & 75.89$\pm$0.30 & 75.74$\pm$0.44 & 0.808$\pm$0.045 & 0.830$\pm$0.049 & 5180$\pm$120 & 5220$\pm$130 \\
409970312 & 4315 &    & 133.95$\pm$0.71 & 134.1$\pm$1.3 & 0.922$\pm$0.047 & 0.914$\pm$0.049 & 5770$\pm$130 & 5810$\pm$190 \\
381133463 & 4324 &    & 17.100$\pm$0.025 & 17.038$\pm$0.018 & 0.387$\pm$0.012 & 0.395$\pm$0.012 & 3420$\pm$160 & 3480$\pm$160 \\
332710428 & 4328 &    & 25.220$\pm$0.035 & 25.004$\pm$0.017 & 0.753$\pm$0.053 & 0.740$\pm$0.054 & 4750$\pm$120 & 4710$\pm$130 \\
436933291 & 4336 &    & 22.563$\pm$0.033 & 22.546$\pm$0.038 & 0.3233$\pm$0.0096 & 0.3266$\pm$0.0099 & 3260$\pm$160 & 3370$\pm$160 \\
44454210 & 4347 &    & 195.8$\pm$2.1 & 195.1$\pm$1.1 & 1.534$\pm$0.050 & 1.563$\pm$0.049 & 7580$\pm$140 & 7610$\pm$210 \\
50531031 & 4348 &    & 138.37$\pm$0.86 & 136.79$\pm$0.46 & 1.853$\pm$0.089 & 1.863$\pm$0.087 & 6090$\pm$130 & 6130$\pm$130 \\
165624043 & 4363 &    & 78.05$\pm$0.14 & 77.78$\pm$0.22 & 0.695$\pm$0.070 & 0.691$\pm$0.071 & 4010$\pm$130 & 4000$\pm$120 \\
405545107 & 4364 &    & 43.913$\pm$0.047 & 43.97$\pm$0.10 & 0.463$\pm$0.014 & 0.466$\pm$0.014 & 3520$\pm$160 & 3500$\pm$160 \\
355793606 & 4370 &    & 211.9$\pm$1.8 & 212.0$\pm$2.4 & 1.89$\pm$0.11 & 1.900$\pm$0.090 & 6150$\pm$150 & 6160$\pm$140 \\
29959419 & 4382 &    & 164.80$\pm$0.98 & 164.4$\pm$1.3 & 1.459$\pm$0.082 & 1.455$\pm$0.067 & 6260$\pm$160 & 6260$\pm$140 \\
454186037 & 4403 &    & 339.6$\pm$5.1 & 339.3$\pm$2.2 & 1.726$\pm$0.092 & 1.724$\pm$0.083 & 5570$\pm$130 & 5520$\pm$120 \\
311279861 & 4422 &    & 360.4$\pm$6.7 & 359.6$\pm$8.3 & 1.856$\pm$0.064 & 1.830$\pm$0.060 & 8020$\pm$100 & 8000$\pm$140 \\
91694663 & 4459 &    & 255.1$\pm$3.0 & 255.0$\pm$2.1 & 0.577$\pm$0.018 & 0.577$\pm$0.017 & 3740$\pm$160 & 3750$\pm$160 \\
158816603 & 4463 &    & 173.5$\pm$1.4 & 173.1$\pm$1.3 & 1.080$\pm$0.056 & 1.090$\pm$0.060 & 5580$\pm$140 & 5540$\pm$140 \\
426389001 & 4602 &    & 62.54$\pm$0.20 & 63.06$\pm$0.20 & 1.151$\pm$0.054 & 1.157$\pm$0.053 & 6040$\pm$120 & 6010$\pm$120 \\
270345987 & 4606 &    & 134.19$\pm$0.86 & 130.72$\pm$0.76 & 1.232$\pm$0.060 & 1.220$\pm$0.060 & 6000$\pm$130 & 5980$\pm$140 \\
331606386 & 4607 &    & 175.8$\pm$1.0 & 180.0$\pm$1.5 & 1.324$\pm$0.057 & 1.310$\pm$0.060 & 6480$\pm$240 & 6400$\pm$160 \\
249064742 & 4610 &    & 48.60$\pm$0.11 & 47.91$\pm$0.12 & 0.666$\pm$0.063 & 0.690$\pm$0.070 & 4160$\pm$120 & 4100$\pm$120 \\
238858993 & 4804 &    & 500$\pm$14 & 495.0$\pm$8.4 & 2.57$\pm$0.15 & 2.62$\pm$0.12 & 6530$\pm$110 & 6680$\pm$150 \\
333660029 & 5749 &    & 614$\pm$15 & 609.0$\pm$7.6 & 3.54$\pm$0.21 & 3.40$\pm$0.17 & 5960$\pm$140 & 6020$\pm$130 \\
397369897 & 5838 &    & 216.9$\pm$1.4 & 218.1$\pm$1.5 & 0.737$\pm$0.061 & 0.750$\pm$0.060 & 4510$\pm$130 & 4460$\pm$140 \\
22018578 & 5856 &    & 302.0$\pm$2.4 & 298.0$\pm$3.7 & 1.293$\pm$0.072 & 1.350$\pm$0.070 & 5900$\pm$140 & 5810$\pm$150 \\
116279757 & 6005 &    & 111.76$\pm$0.29 & 107.80$\pm$0.22 & 0.688$\pm$0.041 & 0.733$\pm$0.043 & 4960$\pm$120 & 4820$\pm$110 \\
\enddata
\end{deluxetable*}

\end{document}